\documentclass[12pt]{article}

\usepackage{scicite}

\usepackage{times}
\usepackage{braket}

\usepackage{graphicx}
\usepackage{caption}

\newenvironment{sciabstract}{%
\begin{quote} \bf}
{\end{quote}}

\newcounter{lastnote}

\title{New material platform for superconducting transmon qubits with coherence times exceeding 0.3 milliseconds}

\author
{Alex P. M. Place$^{1\ast}$, Lila V. H. Rodgers$^{1\ast}$, Pranav Mundada$^{1}$,\\ 
Basil M. Smitham$^{1}$, Mattias Fitzpatrick$^{1}$, Zhaoqi Leng$^{2}$, \\
Anjali Premkumar$^{1}$, Jacob Bryon$^{1}$, Sara Sussman$^{2}$, \\
Guangming Cheng$^{3}$, Trisha Madhavan$^{1}$, Harshvardhan K. Babla$^{1}$, \\
Berthold J{\"a}ck$^{2}$, Andr\'as Gyenis$^{1}$, Nan Yao$^{3}$, \\
Robert J. Cava$^{4}$, Nathalie P. de Leon$^{1}$, Andrew A. Houck$^{1\dagger}$ \\
\\
\small{$^{1}$Department of Electrical Engineering, Princeton University, Princeton, NJ 08544, USA.}\\
\small{$^{2}$Department of Physics, Princeton University, Princeton, NJ 08544, USA.}\\
\small{$^{3}$Princeton Institute for Science and Technology of Materials,}\\
\small{Princeton University, Princeton, NJ 08544, USA.}\\
\small{$^{4}$Department of Chemistry, Princeton University, Princeton, NJ 08544, USA.} \\
\\
\small{$^{\ast}$These authors contributed equally to this work.} \\
\small{$^\dagger$To whom correspondence should be addressed; E-mail: aahouck@princeton.edu.}}

\date{}

\begin{document} 


\baselineskip24pt


\maketitle


\begin{sciabstract}
The superconducting transmon qubit is a leading platform for quantum computing and quantum science. Building large, useful quantum systems based on transmon qubits will require significant improvements in qubit relaxation and coherence times, which are orders of magnitude shorter than limits imposed by bulk properties of the constituent materials. This indicates that relaxation likely originates from uncontrolled surfaces, interfaces, and contaminants. Previous efforts to improve qubit lifetimes have focused primarily on designs that minimize contributions from surfaces. However, significant improvements in the lifetime of two-dimensional transmon qubits have remained elusive for several years. Here, we fabricate two-dimensional transmon qubits that have both lifetimes and coherence times with dynamical decoupling exceeding 0.3 milliseconds by replacing niobium with tantalum in the device. We have observed increased lifetimes for seventeen devices, indicating that these material improvements are robust, paving the way for higher gate fidelities in multi-qubit processors.

\end{sciabstract}

% In setting up this template for *Science* papers, we've used both
% the \section* command and the \paragraph* command for topical
% divisions.  Which you use will of course depend on the type of paper
% you're writing.  Review Articles tend to have displayed headings, for
% which \section* is more appropriate; Research Articles, when they have
% formal topical divisions at all, tend to signal them with bold text
% that runs into the paragraph, for which \paragraph* is the right
% choice.  Either way, use the asterisk (*) modifier, as shown, to

Steady progress in improving gate fidelities for superconducting qubits over the last two decades has enabled key demonstrations of quantum algorithms \cite{QuantumAlgorithmReview,kandala2017hardware,o2016scalable}, quantum error correction \cite{ErrorCorrectionDemonstration,hu2019quantum,reed2012realization}, and quantum supremacy \cite{quantumSupremacy}. These demonstrations have relied on either improving coherence through microwave engineering to avoid losses associated with surfaces and interfaces \cite{paik3D,oliver2013materials,gambetta2016investigating} and to minimize the effects of thermal noise and quasiparticles \cite{potQuasiparticles,serniak2018hot,gustavsson2016suppressing,wang2014measurement}, or by realizing fast gates using tunable coupling \cite{chen2014qubit,mundada2019suppression}. By contrast, little progress has been made in addressing the microscopic source of loss and noise in the constituent materials. Specifically, the lifetime ($T_1$) of the two-dimensional (2D) transmon qubit has not reliably improved beyond 100 $\mu$s since 2012 \cite{kjaergaard2019superconducting,devoretOverview}, and to date the longest published $T_1$ is 114 $\mu$s \cite{rigettiFab}, consistent with other recent literature reports \cite{errorCorrectionThreshold,Wei_top2DIBM,dunsworth_top2DGoogle}.

The lifetimes of current 2D transmons are believed to be limited by microwave dielectric losses \cite{annealOliver,lisenfeld2019electric,wang2015surface}. However, the expected loss tangent of the bulk constituent materials should allow for significantly longer lifetimes. For example, high-purity bulk sapphire has a loss tangent less than $10^{-9}$ \cite{braginsky1987experimental,luiten1993ultrahigh}, which would enable $T_1$ to exceed 30 ms. This suggests that losses are dominated instead by uncontrolled defects at surfaces and interfaces, or by material contaminants. Here we demonstrate that a significant improvement over the state of the art in 2D transmon qubits can be achieved by using tantalum as the superconductor in the capacitor and microwave resonators, replacing the more commonly used niobium. We hypothesize that the complicated stoichiometry of oxides at the niobium surface \cite{cavaNboxidePRB, nico2016niobium} leads to additional microwave loss, and that the insulating oxide of tantalum \cite{face1987nucleation, face1986Taoxide} reduces microwave loss in the device. We observe a time-averaged $T_1$ exceeding 0.3 ms  in our best device and an average $T_1$ of 0.23 ms averaged across all devices, a significant improvement over the state of the art.  

To fabricate qubits \cite{materialsAndMethods}, we deposit tantalum on sapphire substrates by sputtering while heating the substrate to around 500$^{\circ}$C to ensure growth of the $\alpha$ phase \cite{taHotdepAlphaphase,proberThesis}. We then use photolithography and a wet chemical etch to define the capacitor and resonator of the device, followed by electron beam lithography and electron beam evaporation of aluminum and aluminum oxide to form Josephson junctions (Fig. 1A). Between most key steps of the fabrication process, we use solvent and piranha cleaning to reduce contamination introduced during fabrication. The transmon is capacitively coupled to a lithographically-defined cavity (Fig. 1B), allowing us to dispersively measure the state of the qubit \cite{wallraff2005approaching}. To determine $T_1$, we excite the qubit with a $\pi$-pulse and measure its decay over time at a temperature between 9 and 20 mK. In our best device, we measure a peak $T_1$ of 0.36 $\pm$ 0.01 ms (Fig. 1C). We verify that the deposited tantalum film is in the BCC $\alpha$ phase by measuring resistance as a function of temperature. The observed superconducting critical temperature ($T_c$) is around 4.3 K, which is consistent with the intended phase (Fig. 1D) rather than the tetragonal $\beta$ phase which has a $T_c$ below 1K \cite{proberTaTc,readBetaTa}. 

We observe reproducible, robust enhancement of $T_1$ across all devices fabricated with this process. The lifetime of a given qubit fluctuates over time, with a standard deviation of around $7\%$ of the mean (Fig. 2A). Results for eight devices are presented in Fig. 2B, with the time-averaged $T_1$ ranging from 0.15 ms to 0.30 ms, and an average $T_1$ of 0.23 ms across all devices, qualitatively exceeding the $T_1$ of prior 2D transmon devices. The time-averaged coherence time, $T_{2,Echo}$, in our best device is 0.20 $\pm$ 0.03 ms (a trace is shown in Fig. 2C). We can extend the coherence time using a Carr-Purcell-Meiboom-Gill (CPMG) pulse sequence \cite{bylander2011noise} (Fig. 2D), and we achieve a time-averaged $T_{2,CPMG}$ of 0.38 $\pm$ 0.11 ms in our best device (Fig. 2A). The spectral noise density extracted from dynamical decoupling measurements is consistent with $1/f$ noise (Fig. S12).

In addition to the eight qubits presented in Fig. 2, we present data on a total of 23 transmon qubits that were fabricated using different geometries, materials and fabrication processes in the supplementary text (Table S1). We note that switching from niobium to tantalum alone increased the average $T_1$ for a Purcell-filtered qubit to 150 $\mu$s (Device 2a), already a significant improvement over the best published 2D transmon lifetime. To study the impact of heating the substrate during deposition, we made a device from niobium sputtered at 500$^{\circ}$C (Device Nb2). This resulted in a $T_1$ of 79 $\pm$ 1 $\mu$s, an improvement over our previous niobium devices, but not comparable to tantalum-based devices. This indicates that thermal cleaning of the substrate may play a role in enhancing $T_1$, but does not completely explain our improved coherence.

Iterative improvements to processing, including the use of wet etching to pattern the tantalum layer and the introduction of additional cleaning steps, further improved qubit lifetimes to the levels reported in Fig. 2 (Devices 11-18). Specifically, a piranha cleaning process\cite{materialsAndMethods} was introduced to clean particulates and contaminants from the substrate surface. Removal of particulates can be verified using atomic force microscopy (AFM), and the signal due to adventitious carbon measured by x-ray photoelectron spectroscopy (XPS) is attenuated after cleaning. In addition, we find that the introduction of an optimized wet etch process to pattern the tantalum resulted in improved edge morphology compared with reactive ion etching. Of the ten devices measured prior to the optimized wet etch, none had a $T_1$ in excess of 200 $\mu$s; of the eight patterned with the optimized wet etch and fabricated with our substrate cleaning procedure, six had a $T_1$ greater than 200 $\mu$s.  We speculate that metal residue and poor edge morphology may limit qubit lifetimes \cite{materialsAndMethods}. 

Because the crystal structure of thin tantalum films sensitively depends on deposition parameters\cite{taHotdepAlphaphase,proberTaTc}, we present detailed characterization of the deposited tantalum films. Microscopy and spectroscopy of the deposited tantalum confirms the BCC structure of the film and reveals that it is highly oriented. Scanning transmission electron microscopy (STEM) of a film cross section reveals a columnar structure, with the growth direction oriented along the $[110]$ axis (Fig. 3A). We confirm that the films are oriented over a larger area using x-ray diffraction (XRD) measurements \cite{materialsAndMethods}. Atomic-resolution STEM reveals that the individual columnar grains are single-crystal, with the front growth face perpendicular to either the $\braket{100}$ or $\braket{111}$ directions (Fig. 3B). The different orientations result from the underlying three-fold symmetry of the sapphire c-plane surface \cite{oshima2007ga2o3}. A top-down plane view cross-sectional STEM shows that the grains range in size from around 5 to 50 nm (Fig. 3C). Elemental analysis using energy dispersive spectroscopy (EDS) shows that there is no oxide growth between the grains, and the image contrast observed in Fig. 3C arises from diffraction contrast due to interfacial defects at grain boundaries \cite{materialsAndMethods}. 

We also study properties of the native tantalum oxide in our devices using photoelectron spectroscopy to probe large areas and electron microscopy to directly image the oxide layer in a small cross section. XPS shows a set of four peaks with binding energy between 20 and 30 eV, assigned to the Ta 4f core ionization. The two lower binding energy peaks are spin-orbit split peaks associated with Ta metal, while the two higher binding energy peaks are consistent with Ta$_2$O$_5$ (Fig. 3D) \cite{mcguireCoreBETa, himpselTaXPSshiftPRB}. The small peaks at higher binding energies likely correspond to 5p photoelectron emission from the metal and oxide \cite{moulder1995handbook}. The relative intensity of the Ta and Ta$_2$O$_5$ peaks indicates that the oxide is approximately 2 nm thick, given an inelastic mean free path of electrons in tantalum of 2 nm at 1480 eV \cite{TaIMFP}.  This is consistent with measurements of the oxide thickness using angle-resolved XPS \cite{materialsAndMethods, TaIMFP}. High-resolution STEM also verifies that there is a 2-3 nm thick amorphous layer at the surface of the film \cite{materialsAndMethods}. We observe that the apparent oxide thickness and composition are similar across different depositions and are robust to processing steps, including lithography and piranha etching \cite{materialsAndMethods}.

We directly image the interface between the sapphire surface and the sputtered tantalum using integrated differential phase contrast imaging (iDPC) under STEM (Fig. 3E). The interface shows an atomically sharp boundary with clear evidence of epitaxial growth, in which the tantalum atomic layer is directly grown on top of the oxygen atomic layer in the sapphire. The interfacial dislocations likely result from the 12.6$\%$ lattice mismatch between the [$\bar{1}$12] axis of tantalum and the [11$\bar{2}$0] axis of sapphire \cite{materialsAndMethods}, as well as atomic layer steps in the sapphire that are evident in the STEM image.

We have demonstrated that tantalum 2D transmon qubits exhibit longer $T_1$ and $T_2$ than the previous state of the art with remarkable consistency. Building on these relatively simple materials improvements, there are several areas of future exploration. First, $T_{2,Echo}$ is shorter than $T_1$ for all tantalum devices measured. Combining our devices with recent improvements in shielding \cite{kreikebaum2016optimization} and filtering \cite{wang2019cavity} will allow us to explore the microscopic mechanisms for decoherence. Additionally, ongoing work includes more systematic characterization of the effects of specific material properties on microwave losses. In particular, we are exploring the impact of tantalum grain size, oxide thickness, and heteroepitaxial growth interface quality on $T_1$ and $T_2$. Furthermore, it has been well-established that multi-qubit devices suffer from significant variation between qubits \cite{Wei_top2DIBM}, as well as variation over time in the same qubit \cite{klimov2018fluctuations}. An interesting question is how particular material choices quantitatively affect these variations, and whether judicious material choice can narrow the distribution of device properties. Finally, although we have observed a large improvement over niobium-based devices, we note that several groups employ all-aluminum qubits \cite{dunsworth_top2DGoogle,paik3D}. We plan to explore the role of contamination and interfaces in the coherence of all-aluminum qubits, as well as the possibility of fabricating all-tantalum qubits.

More broadly, our results demonstrate that systematic materials improvements are a powerful approach for rapid progress in improving quantum devices. We have recently employed similar targeted materials techniques to improve spin coherence of shallow nitrogen vacancy centers in diamond \cite{sangtawesin2019origins}, and we note that many other quantum platforms are also limited by noise and loss at surfaces and interfaces, including trapped ions \cite{stick2006ion, hiteIonTrapSurface}, shallow donors \cite{schenkel2006electrical, paikIonTrapShallowDonor}, and semiconductor quantum dots \cite{yoneda2018quantum}. Our general approach may allow for directed, rational improvements in these broad classes of systems as well.

\section*{Acknowledgements}
This work was supported by the Army Research Office (HIPS W911NF1910016), the National Science Foundation (MRSEC DMR-1420541 and RAISE DMR-1839199), and the Air Force Office of Scientific Research (FA9550-18-1-0334). A.P.M.P. and L.V.H.R. were supported by the National Defense Science and Engineering Graduate Fellowship. M.F. was supported by an appointment to the Intelligence Community Postdoctoral Research Fellowship Program at Princeton University by the Oak Ridge Institute for Science and Education (ORISE) through an interagency agreement between the U.S. Department of Energy and the Office of the Director of National Intelligence (ODNI). A.P. acknowledges support from the National Science Foundation Graduate Research Fellowship Program. B.J. acknowledges support from a postdoctoral fellowship through the Alexander-von-Humboldt foundation. We thank Paul Cole, Eric Mills, Roman Akhmechet, Bert Harrop, and Joe Palmer for useful discussions relating to fabrication. We also acknowledge help from Daniel Gregory, John Schrieber, and Austin Ferrenti regarding materials characterization. We thank Jeff Thompson for helpful discussions. Devices were fabricated in the Princeton University Micro/Nano Fabrication Laboratory and the Quantum Device Nanofabrication Laboratory. Microscopy and analysis relied on Princeton’s Imaging and Analysis Center.

\section{Supplemental Information}

\subsection{Fabrication procedure}
The 2D transmon qubits are fabricated on c-plane sapphire substrates (Crystec GmbH) that are 0.53 mm thick and double-side polished (Fig. S1). Prior to deposition, the wafer is dipped in a piranha solution then cleaned with an oxygen plasma (Technics PE-IIA System) immediately before loading into the sputterer.

Tantalum is deposited on the sapphire substrate at high temperature (approximately 500$^{\circ}$C, Star Cryoelectronics). Before photolithography, the tantalum-coated substrates are placed in a 2:1 mixture of H$_2$SO$_4$ and H$_2$O$_2$ for 20 minutes (hereafter "piranha-cleaned”) then heated on a hotplate for 5 minutes at 140$^{\circ}$C before AZ 1518 resist is spun (Merck KGaA). The resist is patterned using a direct-write process (2 mm write head on a Heidelberg DWL 66+ Laser Writer). After developing (85 sec in AZ 300MIF developer from Merck KGaA), the resist is hard-baked for 2 min at 115$^{\circ}$C. Unwanted residual resist is removed using a gentle oxygen descum (2 min in 30 mTorr O$_2$ with 20 W/200 W RF/ICP coil power in a Plasma-Therm Apex SLR). Next, the tantalum is etched in a 1:1:1 ratio of HF:HNO$_3$:H$_2$O (Tantalum Etchant 111 from Transene Company, Inc.) for 21 sec. After stripping resist, the device is solvent-cleaned by sonicating in toluene, acetone, methanol, and isopropyl alcohol ("TAMI-cleaned”) then piranha-cleaned. The patterned tantalum is prepared for electron beam lithography to define Josephson junctions (MMA 8.5 MAA, 950 PMMA, with a 40 nm layer of evaporated aluminum to dissipate charge), then the chips are diced into 7x7 mm squares. 

Liftoff patterns for Manhattan junctions \cite{potts2001cmos} with overlap areas of approximately 0.03 $\mu$m$^2$ are then exposed (Elionix ELS-F125). The anticharge layer is removed through a 4 min bath in MF 319 (Rohm and Haas Electronic Materials LLC) followed by a 50 sec bath in a 1:3 mixture of methyl isobutyl ketone to isopropyl alcohol. Next, the device is loaded into a Plassys MEB 550S electron beam evaporator and ion-milled (400 V, 30 sec along each trench of the junction). Immediately after, 15 nm of aluminum is deposited at 0.4 nm/sec at a pressure of approximately 10$^{-7}$ mBar, followed by a 15 min, 200 mBar oxidation period. Finally, 54 nm of aluminum is deposited to form the second layer of the junction, with the same evaporation parameters (for Device 18a, 15 nm and 19 nm of aluminum are deposited, respectively). The resist is then removed by soaking the sample in Remover PG (Kayaku Advanced Materials, Inc.) for approximately 3 hours at 80$^\circ$C, briefly sonicating in hot Remover PG, then swirling in isopropyl alcohol.

\subsubsection{Tantalum Etch} 
Initially we etched tantalum using a reactive-ion etch (8:3:2 CHF$_3$:SF$_6$:Ar chemistry at 50 mTorr, RF/ICP power of 100 W/100 W). However, scanning electron microscopy (SEM) images showed rough edges as well as small pillars and boulders near the sidewalls, likely due to micromasking (Fig. S2A, B). The anomalous objects in Fig. S2B remained after the device was cleaned in piranha solution and treated in an oxygen plasma. In order to avoid these fabrication problems, we employed a wet etch composed of 1:1:1 HF:HNO$_3$:H$_2$O. We found that several resists delaminated before the tantalum was etched through, leaving the sidewalls and nearby tantalum visibly rough in SEM (Fig. S2C). This problem was circumvented by using thick AZ 1518 resist (approximately 2 $\mu$m tall) which left cleaner sidewalls (Fig. S2D). Comparing Devices 4-10 with Devices 11-18 in Table S1, we note the optimized wet etch likely improved $T_1$.

\subsubsection{Sapphire Preparation}

We use aggressive cleans and etches to remove the contaminants on our samples after dicing, stripping resist, and sonicating in solvents. In particular, AFM reveals an abundance of particulates on the surface (Fig. S3A), which are removed by cleaning in piranha solution (Fig. S3B). Additionally, the carbon signal in XPS is attenuated by a factor of 5 after piranha cleaning, illustrating a reduction in carbon contamination (Fig. S3C). XPS also reveals zinc contamination that persists through a piranha clean, but can be removed by etching the sapphire substrate in heated sulfuric acid (Fig. S3D).

We prepare the sapphire surface using this sulfuric acid etch in Devices 9-14 and 17. In these devices, the wafers are covered with a protective layer of photoresist and then diced into 1 inch squares. After removing resist, the squares are TAMI cleaned and piranha cleaned. Next, the sapphire is placed into a quartz beaker filled with H$_2$SO$_4$ sitting on a room temperature hotplate. The hotplate is set to 150$^{\circ}$C for 20 minutes, followed by a 10 minute cooldown period before removing the device. We estimate less than 1 nm of the surface is removed through this procedure \cite{sapphireProcessing}. To avoid residue from the etch, the device is piranha cleaned again. The device is then packaged, shipped, and loaded into a sputterer without further cleaning.

Calibrating the time and temperature of the sapphire etch is critical to maintaining a smooth surface morphology while still removing zinc. In particular, polycrystalline aluminum sulfates form on the sapphire surface after heating in sulfuric acid for too long and at too high of a temperature (Fig. S4A) \cite{sapphireProcessing}. We developed our sapphire etch recipe by $(1)$ looking for crystal formation in an optical microscope, $(2)$ ensuring that zinc was removed in XPS, and $(3)$ checking that we preserved smooth surface morphology in AFM. We note that the zinc appeared to be inhomogeneously distributed on the surface and so we routinely checked multiple spots in XPS.  After adjusting the time and temperature to the optimum procedure outlined above, we did not detect any crystal formation.

Additionally, we observed surface contamination with AFM from etching sapphire in borosilicate glassware. An example of surface particulate contamination is shown in Fig. S4B. Switching to a quartz beaker solved this issue.

We note that Devices 16 and 18 were not processed using the sapphire etch, and they exhibited $T_1$ over 0.2 ms. In the future we are interested in studying the impact of sapphire material properties on device performance. We plan to fabricate devices on higher-purity sapphire, remove polishing-induced strain by etching a more appreciable amount of the substrate, and anneal to form an atomically smooth surface \cite{sapphireProcessing}. 

\subsection{Device Packaging}
The completed devices are first mounted to a printed circuit board (PCB). The edge of the tantalum ground plane is firmly pressed against the PCB's copper backside, sandwiched between the PCB and a piece of aluminum-coated oxygen-free copper. The device is then wirebonded (Fig. S5A, B). An aluminum-coated oxygen-free copper lid is sometimes placed above the qubit (Table S1 column "Enclosure Lid Removed"), forming a superconducting enclosure partially surrounding the qubit. The device is mounted in a dilution refrigerator with a base temperature of approximately 9-20 mK. The qubit and PCB are wrapped in several layers of aluminized mylar sheeting and suspended by an oxygen-free copper rod in the middle of an aluminum cylinder coated with microwave-attenuating epoxy or sheeting (Laird Performance Materials Eccosorb Cr or Loctite Stycast). This cylinder is enclosed in a mu-metal can to reduce the penetration of ambient magnetic fields into the aluminum during the superconducting transition. Both cans are then wrapped in several layers of mylar sheeting. 

We note that all of the double-pad transmons presented in this text are positioned approximately 2 mm away from the copper traces on the PCB (Fig. S5A), which could result in loss due to parasitic coupling of the qubit to the resistive traces. In order to reduce this possible source of loss, devices fabricated with the Xmon geometry were moved close to the center of the sapphire chip (Fig. S5B).
  
\subsection{Measurement Setup}
Each transmon is capacitively coupled to a microwave resonator, allowing the state of the qubit to be measured dispersively \cite{wallraff2005approaching}. The transmon frequencies range from 3.1 - 5.5 GHz while the resonators range in frequency from 6.8 - 7.3 GHz. An overview of the setup used to measure a majority of the devices is given in Fig. S6. An Agilent E8267D vector signal generator, Holzworth HS9004A RF synthesizer, and Keysight M9330A Arbitrary Waveform Generator are used to synthesize the excitation and measurement pulses. The input signals are combined into a single line and then attenuated on each plate of the dilution refrigerator. An additional filter made of Eccosorb CR110 epoxy is placed in the aluminum can to attenuate high-frequency radiation. Measured in reflection, the output signal is sent through a circulator (Raditek RADC-4-8-cryo-0.01-4K-S23-1WR-ss-Cu-b), two isolators (Quinstar QCI -075900XM00), superconducting wires, and then a high-electron-mobility transistor amplifier (Low Noise Factory LNF-LNC4\textunderscore8C) at 4 K. After the signal is amplified at room temperature (through two MITEQ AFS4-00101200 18-10P-4 amplifiers), it is measured in a homodyne setup by first mixing it with a local oscillator (Holzworth HS9004A), further amplifying (Stanford Research Systems  SR445a), and then digitizing (Acqiris U1084A).

\subsection{Other Device Designs and Fabrication Processes}
Table S1 summarizes different iterations of the fabrication procedure. Initially we made a tantalum transmon using our standard niobium processing techniques (reactive ion etching, no acid cleaning). This material switch alone improved the coherence time by more than a factor of four compared to the control sample (Table S1, Devices 1a and Nb1). We then began to iterate our packaging and fabrication techniques to explore the new dominant loss mechanisms.

First we minimized losses unrelated to the qubit materials and interfaces. We reduced the density of photonic states at the qubit frequency by means of a Purcell filter (Device 2a and all subsequent devices) \cite{reed2010fast}. We also deposited aluminum shielding on a majority of the copper enclosure immediately surrounding the device to reduce dissipative currents induced by the qubit in the surrounding metal. At the same time, we introduced a mylar sheet wrapped around the PCB as an extra layer of shielding. Both added layers give additional protection from high-energy radiation (Device 2b and all subsequent devices). 

Next we focused on reducing material contaminants. XPS measurements revealed significant carbon residue that persisted after solvent-based cleaning. Accordingly, we reduced carbon contamination by adding a piranha clean before spinning e-beam resist (Device 4a and all subsequent devices). As mentioned above, we also cleaned the sapphire substrate prior to tantalum deposition. For Devices 1-8, 15-16, and 18 as well as Nb1, Nb2, and 3D1, the sapphire substrate was dipped in a piranha solution and cleaned with an oxygen plasma (Technics PE-IIA System) immediately before loading into the sputterer. For the rest of the sapphire devices, we cleaned the substrate with the sapphire etch described above (Section S1.1), packaged and shipped the samples, then deposited the tantalum.

We then focused on the tantalum etch, described in more detail above (Section S1.1). Devices 1-6, Nb 1-2, Si1, and 3D1 were all fabricated with reactive ion etching. Devices 7-10 and 3D2 were made using initial versions of the wet etch (using different resists, etch times, and acid concentrations), where the etch clearly roughened the sidewalls (Fig. S2C). Devices 11-18 were made using the optimized wet etch.

\subsubsection{2D Transmons on Sapphire}
We measured two different geometries of transmons: devices with double-pad capacitors, where neither pad has a direct ground connection (Fig. 1A),  and Xmon-style devices \cite{barends2013coherent}, where the ground plane serves as one side of the transmon's capacitor.

\subsubsection{2D Transmon on Silicon}
We fabricated a 2D, double-pad, tantalum transmon on silicon (Device Si1) with a similar design to that used for the devices on sapphire. The primary elements that changed during the fabrication process were: (i) a different plasma etch time to avoid overetching into the silicon, (ii) no aluminum layer was deposited on top of the e-beam resist prior to e-beam lithography, and (iii) the e-beam intensity was adjusted during the lithography step. We found that reactive-ion etching severely roughened the silicon surface (17 nm RMS surface roughness, measured with a Keyence Optical Profilometer). We plan to optimize this fabrication process in the future.

\subsubsection{3D Transmons}
We mounted our 3D transmons in an 8.0 GHz aluminum rectangular cavity with a 250 kHz linewidth \cite{paik3D}. Double-pad transmons were fabricated with the same process described above on approximately 2.5 $\times$ 7.5 mm sapphire chips. After the aluminum cavity was etched (4 hours in Aluminum Etchant Type A from Transene Company, Inc.) to remove any contaminants and machining cracks, the device was mounted in the center of the resonator and indium foil was compressed between the two aluminum halves to seal the seam. The same shielding was used for the 2D and 3D devices other than the aluminum cylinder directly inside the mu-metal shielding, which was too small to fit around the 3D cavity. 

For Device 3D1, our measured cavity resonances were significantly different than expected. We attribute this to a thin layer of aluminum that was deposited on the side of the sapphire chip during the double-angle Josephson junction evaporation. On later devices, we cleaned the metal from the side of the chip and our measured resonance was as expected. We measured a mean $T_1$ of 0.20 $\pm$ 0.02 ms for Device 3D1 (Fig. S7A). We plan to optimize 3D transmon devices in future work, in particular focusing on measurement of 3D devices that were fabricated with wet etch processing.

\subsection{Tantalum Fluxonium}
We fabricated a light fluxonium qubit \cite{manucharyan2009fluxonium} made using tantalum capacitor pads and aluminum junctions. Our qubit has a Josephson energy of 0.92 GHz, a capacitive energy of 3.6 GHz, and an inductive energy of 0.53 GHz. We found a plasmon $T_1$ of 0.063 $\pm$ 0.004 ms and a maximum fluxon $T_1$ of 1.9 $\pm$ 0.2 ms, although we saw time fluctuations in the fluxon $T_1$ on the order of a millisecond. The resonant frequency of a fluxonium qubit is flux-tunable. By fitting $T_1$ as a function of resonant frequency, we deduced a dielectric loss tangent of 1-3$\cdot$10$^{-6}$ \cite{nguyen2019high} (Fig. S7C, D).

\subsection{Additional Materials Characterization}
\subsubsection{X-ray Diffraction}
We use XRD to study the crystal structure of our films over a much larger area than is feasible with STEM images (Fig. S8). An acquired spectrum of a film exhibits a strong peak corresponding to $\alpha$-tantalum [110] \cite{taHotdepAlphaphase}, corroborating STEM images that suggest that our films grow uniformly along that direction (Fig. 3A). Additionally, we observe peaks corresponding to sapphire [006] \cite{sapphireProcessing} and $\alpha$-tantalum [220] \cite{taHotdepAlphaphase}. We do not detect a $\beta$-tantalum [002] peak at 33.7$^{\circ}$ (2$\theta$) \cite{taHotdepAlphaphase}. This provides further evidence along with our $T_c$ and STEM measurements that the tantalum films are uniformly in the $\alpha$ phase. We note that there are a few unassigned small peaks which could result from contamination, instrumental artifacts, or impurities or defects in the tantalum films. 

\subsubsection{Grain Boundaries}
We further interrogate the grain boundaries visible in a plane-view image (Fig. S9A) by using energy dispersive x-ray spectroscopy (EDS) to perform spatially-resolved elemental analysis. We find a uniform distribution of tantalum (Fig. S9B) and oxygen (Fig. S9C) over the region, and no oxygen enrichment at the grain boundaries. This suggests that our films do not grow oxide between the grains, and that the image contrast observed in Fig. S9A arises instead from diffraction contrast caused by interfacial defects. 

A high-resolution STEM image of a grain boundary elucidates the crystal structure at the boundaries (Fig. S9D). Taking a diffraction pattern of a grain boundary region indicated by a green square in Fig. S9D gives a pattern consistent with twinning (Fig. S9E).  A diffraction pattern of the whole region in Fig. S9D illustrates the rotational symmetries of the grains (Fig. S9F).

\subsubsection{Tantalum Oxide}
An atomic-resolution STEM image of a 50 nm region of the tantalum surface reveals an amorphous oxide that is 2-3 nm thick (Fig. S10A). We further study this oxide using XPS to estimate oxide thickness and composition over a larger area (250 $\mu$m spot size) (Fig. S10B, D, E, F). XPS scans of the tantalum film show two sharp lower binding energy peaks assigned to tantalum metal 4f$_{7/2}$ and 4f$_{5/2}$ orbitals (lower binding energy to higher binding energy, respectively), two peaks at higher binding energy corresponding to the same orbitals of Ta$_2$O$_5$ \cite{himpselTaXPSshiftPRB, mcguireCoreBETa}, and two small 5p$_{3/2}$ peaks corresponding to the metal and oxide, respectively \cite{moulder1995handbook}. Assuming the mean free path of electrons in tantalum is 2 nm at 1480 eV \cite{TaIMFP}, and only taking into account inelastic scattering, a thickness can be estimated by comparing the ratio of oxide to metal peak areas. We corroborate this estimation using angle-resolved XPS (ARXPS), where we vary the angle between sample and detector, changing the relative distances that the emitted photoelectrons travel through the metal and oxide layers to reach the detector (Fig. S10B). We account for this geometry in our modeling, and extract the oxide thickness at different angles (Fig. S10C). The thickness estimation remains fairly consistent until higher angles, when other effects related to surface morphology or elastic scattering become more significant (Fig. S10C) \cite{cumpson1995ARXPS}. 

To investigate the variability of oxide thickness between devices, we show normal incidence XPS data from three devices from different tantalum depositions with different surface cleaning fabrication procedures (Fig. S10D, E, F). In addition to variations in other fabrication steps, we  note that the device shown in Fig. S10D was only solvent cleaned, and the devices in Fig. S10E and Fig. S10F were piranha cleaned. The peak shapes and ratio of oxide to metal peak area are highly consistent between all these devices, suggesting there is no appreciable change in oxide thickness or composition. 

\subsubsection{Sapphire-Tantalum Interface}
For completeness, we include an iDPC STEM image showing the interface between sapphire and tantalum viewed from $\braket{1\bar{1}00}$ sapphire and $\braket{100}$ tantalum zone axes (Fig. S11A). We also propose atomistic models for an ideal sapphire-tantalum interface shown in Figs. S11B and C, as a starting point for future studies on the impact of sapphire surface morphology on heteroepitaxial growth.  

\subsubsection{XPS, AFM, XRD characterization}
All XPS, AFM, and XRD data were acquired using tools in the Imaging and Analysis Center at Princeton University. 

XPS was performed using a Thermo Fisher K-Alpha and X-Ray Spectrometer tool with a 250 $\mu$m spot size. The data shown in Fig. 3D, Fig. S3C and D, and Fig. S10D, E, and F were obtained by collecting photoelectrons at normal incidence between sample and detector. The angle-resolved XPS (ARXPS) spectra shown in Fig. S10B were collected by changing the angle between sample and detector. All AFM images were taken with a Bruker Dimension Icon3 tool operating in tapping mode (AFM tip from Oxford Instruments Asylum Research, part number AC160TS-R3, resonance frequency 300 kHz). The XRD spectrum shown in Fig. S6 was collected with a Bruker D8 Discover X-Ray Diffractometer configured with Bragg-Brentano optics. Two 0.6 mm slits were inserted before the sample, and a 0.1 mm slit was placed before the detector. 

\subsubsection{Electron Microscopy Characterization}
SEM and STEM images were also collected at the Imaging and Analysis Center at Princeton University. STEM thin lamellae (thickness: 70-1300 nm) were prepared by focused ion beam cutting via a FEI Helios NanoLab 600 dual beam system (FIB/SEM). All the thin samples for experiments were polished by a 2 keV Ga ion beam to minimize the surface damage caused by the high-energy ion beam. Conventional STEM imaging, iDPC, atomic-resolution HAADF-STEM imaging and atomic-level EDS mapping were performed on a double Cs-corrected Titan Cubed Themis 300 STEM equipped with an X-FEG source operated at 300 kV and a super-X energy dispersive spectrometry (super-X EDS) system.

Lithography and etching process development SEM images were collected with a FEI Verios 460XHR SEM and a FEI Quanta 200 Environmental SEM. Various tilt angles, working distances, and chamber pressures were used to eliminate charging effects. 

\subsection{CPMG}
To reduce our devices' low-frequency noise sensitivity we applied a sequence of $\pi$-pulses \cite{bylander2011noise}. Each pulse had a Gaussian envelope with $\sigma$ around 20-50 ns and was truncated at $\pm2\sigma$.  Due to the large number of sequential pulses, we found that reducing gate error through frequent calibration was important. 

To derive the qubit's noise spectral density (Fig. S12) from such a pulse sequence, we follow the procedure in \cite{bylander2011noise}. Our signal-to-noise ratio is significantly worse, as the overall delay time between initial excitation and measurement increases. For clarity, we include only delays spanning up to approximately $T_1$. For simplicity we assume the gates are instantaneous.  We find a noise power spectral density that is well fit by $A/f^\alpha + B$ with $\alpha$ = 0.7.

\subsection{Fitting Procedure}
We fit our transmon $T_1$ data to $f(\Delta t) = e^{-\Delta t / T_1}$, where $T_1$ is a fit parameter and the function represents the population in the excited state. We fit any $T_2$ data taken with fringes to the fit $f(\Delta t) = 0.5 e^{-\Delta t /T_{2R}} \cos(2\pi \Delta t \delta + \phi_0) + 0.5$ where $T_{2R}$, $\delta$, and $\phi_0$ are fit parameters. For echo and CPMG experiments, we fit our $T_2$ data with a stretched exponential, $f(\Delta t) = 0.5e^{-(\Delta t / T_2)^n} + 0.5$, where $T_2$ and $n$ are fit parameters. If $n < 1$, the data is refit to a pure exponential. Fig. S13 shows a representative decay for a low, average, and high value of $T_{2,CPMG}$ for the data shown in Fig. 2A. In time sequences, data traces with obvious abnormalities or poor fits as measured by root-mean-square error are discarded.
% Your references go at the end of the main text, and before the
% figures.  For this document we've used BibTeX, the .bib file
% scibib.bib, and the .bst file Science.bst.  The package scicite.sty
% was included to format the reference numbers according to *Science*
% style.

\bibliographystyle{Science}
\bibliography{scibib}

\begin{thebibliography}{10}

\bibitem{QuantumAlgorithmReview}
A.~Montanaro, {\it npj Quantum Information\/} {\bf 2}, 15023 (2016).

\bibitem{kandala2017hardware}
A.~Kandala, {\it et~al.\/}, {\it Nature\/} {\bf 549}, 242 (2017).

\bibitem{o2016scalable}
P.~J. O’Malley, {\it et~al.\/}, {\it Physical Review X\/} {\bf 6}, 031007
  (2016).

\bibitem{ErrorCorrectionDemonstration}
N.~Ofek, {\it et~al.\/}, {\it Nature\/} {\bf 536}, 441 (2016).

\bibitem{hu2019quantum}
L.~Hu, {\it et~al.\/}, {\it Nature Physics\/} {\bf 15}, 503 (2019).

\bibitem{reed2012realization}
M.~D. Reed, {\it et~al.\/}, {\it Nature\/} {\bf 482}, 382 (2012).

\bibitem{quantumSupremacy}
F.~Arute, {\it et~al.\/}, {\it Nature\/} {\bf 574}, 505 (2019).

\bibitem{paik3D}
H.~Paik, {\it et~al.\/}, {\it Physical Review Letters\/} {\bf 107}, 240501
  (2011).

\bibitem{oliver2013materials}
W.~D. Oliver, P.~B. Welander, {\it MRS bulletin\/} {\bf 38}, 816 (2013).

\bibitem{gambetta2016investigating}
J.~M. Gambetta, {\it et~al.\/}, {\it IEEE Transactions on Applied
  Superconductivity\/} {\bf 27}, 1 (2016).

\bibitem{potQuasiparticles}
A.~D. C{\'o}rcoles, {\it et~al.\/}, {\it Applied Physics Letters\/} {\bf 99},
  181906 (2011).

\bibitem{serniak2018hot}
K.~Serniak, {\it et~al.\/}, {\it Physical review letters\/} {\bf 121}, 157701
  (2018).

\bibitem{gustavsson2016suppressing}
S.~Gustavsson, {\it et~al.\/}, {\it Science\/} {\bf 354}, 1573 (2016).

\bibitem{wang2014measurement}
C.~Wang, {\it et~al.\/}, {\it Nature communications\/} {\bf 5}, 5836 (2014).

\bibitem{chen2014qubit}
Y.~Chen, {\it et~al.\/}, {\it Physical review letters\/} {\bf 113}, 220502
  (2014).

\bibitem{mundada2019suppression}
P.~Mundada, G.~Zhang, T.~Hazard, A.~Houck, {\it Physical Review Applied\/} {\bf
  12}, 054023 (2019).

\bibitem{kjaergaard2019superconducting}
M.~Kjaergaard, {\it et~al.\/}, {\it arXiv preprint arXiv:1905.13641\/}  (2019).

\bibitem{devoretOverview}
M.~H. Devoret, R.~J. Schoelkopf, {\it Science\/} {\bf 339}, 1169 (2013).

\bibitem{rigettiFab}
A.~Nersisyan, {\it et~al.\/}, {\it arXiv preprint arXiv:1901.08042\/}  (2019).

\bibitem{errorCorrectionThreshold}
R.~Barends, {\it et~al.\/}, {\it Nature\/} {\bf 508}, 500 (2014).

\bibitem{Wei_top2DIBM}
K.~X. Wei, {\it et~al.\/}, {\it arXiv preprint arXiv:1905.05720\/}  (2019).

\bibitem{dunsworth_top2DGoogle}
A.~Dunsworth, {\it et~al.\/}, {\it Applied Physics Letters\/} {\bf 111}, 022601
  (2017).

\bibitem{annealOliver}
A.~Kamal, {\it et~al.\/}, {\it arXiv preprint arXiv:1606.09262\/}  (2016).

\bibitem{lisenfeld2019electric}
J.~Lisenfeld, {\it et~al.\/}, {\it npj Quantum Information\/} {\bf 5}, 1
  (2019).

\bibitem{wang2015surface}
C.~Wang, {\it et~al.\/}, {\it Applied Physics Letters\/} {\bf 107}, 162601
  (2015).

\bibitem{braginsky1987experimental}
V.~Braginsky, V.~Ilchenko, K.~S. Bagdassarov, {\it Physics Letters A\/} {\bf
  120}, 300 (1987).

\bibitem{luiten1993ultrahigh}
A.~Luiten, A.~Mann, D.~Blair, {\it Electronics Letters\/} {\bf 29}, 879 (1993).

\bibitem{cavaNboxidePRB}
R.~J. Cava, {\it et~al.\/}, {\it Physical Review B\/} {\bf 44}, 6973 (1991).

\bibitem{nico2016niobium}
C.~Nico, T.~Monteiro, M.~P. Gra{\c{c}}a, {\it Progress in Materials Science\/}
  {\bf 80}, 1 (2016).

\bibitem{face1987nucleation}
D.~Face, D.~Prober, {\it Journal of Vacuum Science \& Technology A: Vacuum,
  Surfaces, and Films\/} {\bf 5}, 3408 (1987).

\bibitem{face1986Taoxide}
D.~Face, D.~Prober, W.~McGrath, P.~Richards, {\it Applied physics letters\/}
  {\bf 48}, 1098 (1986).

\bibitem{materialsAndMethods}
Materials and methods are available as supplementary materials at the Science
  website.

\bibitem{taHotdepAlphaphase}
L.~Gladczuk, A.~Patel, C.~S. Paur, M.~Sosnowski, {\it Thin Solid Films\/} {\bf
  467}, 150 (2004).

\bibitem{proberThesis}
M.~Reese, Superconducting hot electron bolometers for terahertz sensing, Ph.D.
  thesis, Yale (2006).

\bibitem{wallraff2005approaching}
A.~Wallraff, {\it et~al.\/}, {\it Physical review letters\/} {\bf 95}, 060501
  (2005).

\bibitem{proberTaTc}
D.~Face, D.~Prober, {\it Journal of Vacuum Science \& Technology A: Vacuum,
  Surfaces, and Films\/} {\bf 5}, 3408 (1987).

\bibitem{readBetaTa}
M.~H. Read, C.~Altman, {\it Applied Physics Letters\/} {\bf 7}, 51 (1965).

\bibitem{bylander2011noise}
J.~Bylander, {\it et~al.\/}, {\it Nature Physics\/} {\bf 7}, 565 (2011).

\bibitem{oshima2007ga2o3}
T.~Oshima, T.~Okuno, S.~Fujita, {\it Japanese Journal of Applied Physics\/}
  {\bf 46}, 7217 (2007).

\bibitem{mcguireCoreBETa}
G.~McGuire, G.~K. Schweitzer, T.~A. Carlson, {\it Inorganic Chemistry\/} {\bf
  12}, 2450 (1973).

\bibitem{himpselTaXPSshiftPRB}
F.~Himpsel, J.~Morar, F.~McFeely, R.~Pollak, G.~Hollinger, {\it Physical Review
  B\/} {\bf 30}, 7236 (1984).

\bibitem{moulder1995handbook}
J.~F. Moulder, {\it Physical electronics\/} pp. 170--171 (1995).

\bibitem{TaIMFP}
H.~Shinotsuka, S.~Tanuma, C.~Powell, D.~Penn, {\it Surface and Interface
  Analysis\/} {\bf 47}, 871 (2015).

\bibitem{kreikebaum2016optimization}
J.~M. Kreikebaum, A.~Dove, W.~Livingston, E.~Kim, I.~Siddiqi, {\it
  Superconductor Science and Technology\/} {\bf 29}, 104002 (2016).

\bibitem{wang2019cavity}
Z.~Wang, {\it et~al.\/}, {\it Physical Review Applied\/} {\bf 11}, 014031
  (2019).

\bibitem{klimov2018fluctuations}
P.~Klimov, {\it et~al.\/}, {\it Physical review letters\/} {\bf 121}, 090502
  (2018).

\bibitem{sangtawesin2019origins}
S.~Sangtawesin, {\it et~al.\/}, {\it Physical Review X\/} {\bf 9}, 031052
  (2019).

\bibitem{stick2006ion}
D.~Stick, {\it et~al.\/}, {\it Nature Physics\/} {\bf 2}, 36 (2006).

\bibitem{hiteIonTrapSurface}
D.~Hite, {\it et~al.\/}, {\it MRS bulletin\/} {\bf 38}, 826 (2013).

\bibitem{schenkel2006electrical}
T.~Schenkel, {\it et~al.\/}, {\it Applied Physics Letters\/} {\bf 88}, 112101
  (2006).

\bibitem{paikIonTrapShallowDonor}
S.-Y. Paik, S.-Y. Lee, W.~Baker, D.~McCamey, C.~Boehme, {\it Physical Review
  B\/} {\bf 81}, 075214 (2010).

\bibitem{yoneda2018quantum}
J.~Yoneda, {\it et~al.\/}, {\it Nature nanotechnology\/} {\bf 13}, 102 (2018).

\bibitem{potts2001cmos}
A.~Potts, G.~Parker, J.~Baumberg, P.~de~Groot, {\it IEE Proceedings-Science,
  Measurement and Technology\/} {\bf 148}, 225 (2001).

\bibitem{sapphireProcessing}
F.~Dwikusuma, D.~Saulys, T.~Kuech, {\it Journal of The Electrochemical
  Society\/} {\bf 149}, G603 (2002).

\bibitem{reed2010fast}
M.~D. Reed, {\it et~al.\/}, {\it Applied Physics Letters\/} {\bf 96}, 203110
  (2010).

\bibitem{barends2013coherent}
R.~Barends, {\it et~al.\/}, {\it Physical review letters\/} {\bf 111}, 080502
  (2013).

\bibitem{manucharyan2009fluxonium}
V.~E. Manucharyan, J.~Koch, L.~I. Glazman, M.~H. Devoret, {\it Science\/} {\bf
  326}, 113 (2009).

\bibitem{nguyen2019high}
L.~B. Nguyen, {\it et~al.\/}, {\it Physical Review X\/} {\bf 9}, 041041 (2019).

\bibitem{cumpson1995ARXPS}
P.~J. Cumpson, {\it Journal of Electron Spectroscopy and Related Phenomena\/}
  {\bf 73}, 25 (1995).

\end{thebibliography}

\clearpage

\begin{figure}[!t]
	\begin{center}
		\includegraphics[width=2.25in]{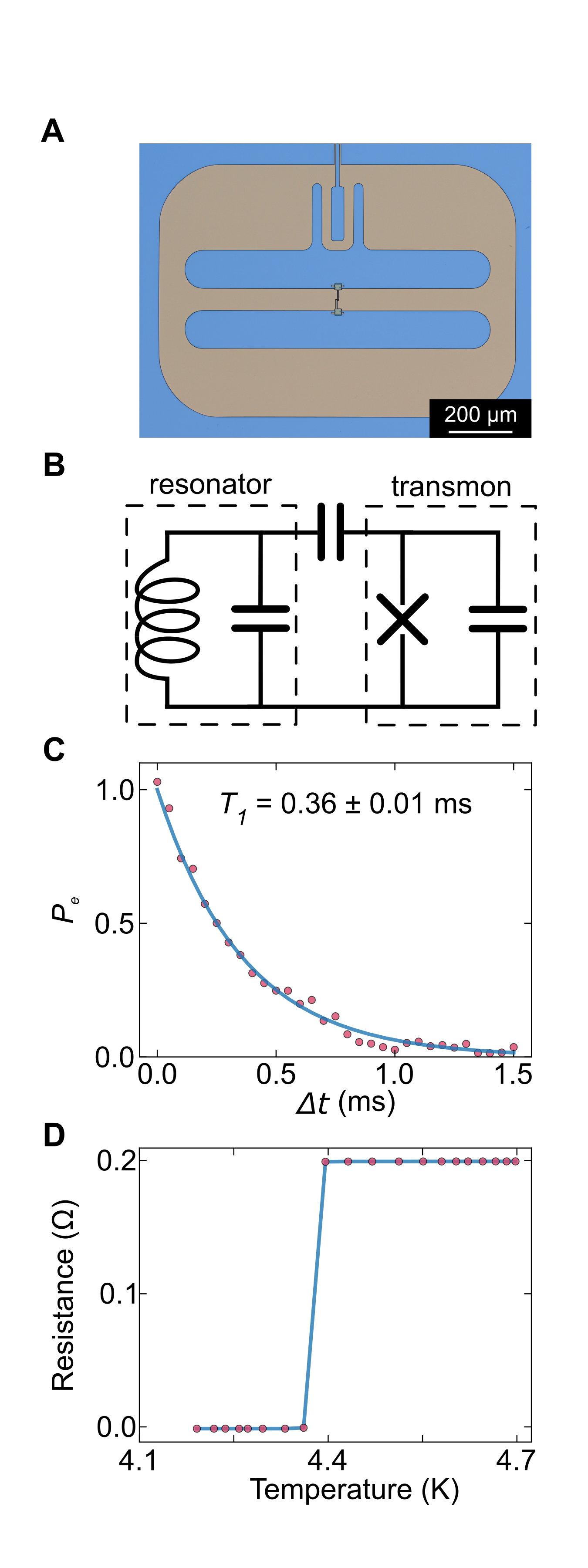}
	\end{center}
	\vspace{-0.6cm}
	\caption*{\textbf{Fig. 1. Tantalum-based transmon superconducting qubit.} (\textbf{A}) False-colored optical microscope image of a transmon qubit. The transmon consists of a Josephson junction shunted by two large capacitor islands made of tantalum (blue). (\textbf{B}) Corresponding circuit diagram of the transmon qubit coupled to the resonator via a coupling capacitor. (\textbf{C}) Peak $T_1$ measurement, showing the excited state population $P_e$ as a function of delay time $\Delta t$. Line represents a single exponential fit with a characteristic $T_1$ time of 0.36 $\pm$ 0.01 ms. (\textbf{D}) Four-probe resistance measurement of the tantalum film showing $T_c$ = 4.38 $\pm$ 0.02 K, consistent with the critical temperature of $\alpha$-tantalum. } 
\end{figure}
%***********

\begin{figure}[!t]
	\begin{center}
		\includegraphics[width=2.25in]{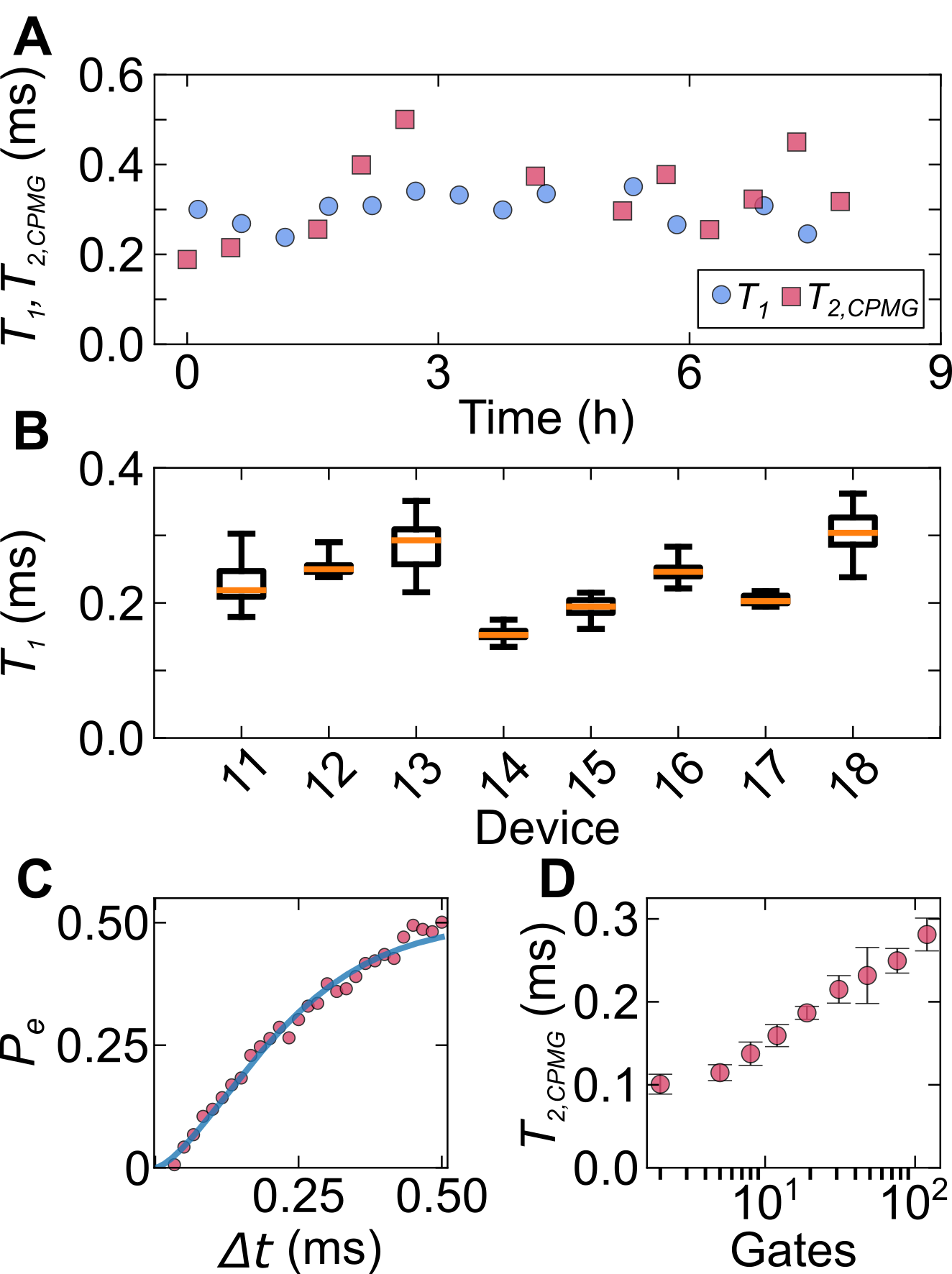}
	\end{center}
	\vspace{-0.6cm}
	\caption*{\textbf{Fig. 2. Lifetime and decoherence measurements.} (\textbf{A}) Measurements of Device 18a over time (fit errors are smaller than the markers). (\textbf{B}) Summary of $T_1$ time series measurements of all devices with optimized processing and packaging. The yellow line shows the median, while the box spans the middle two quartiles of the data. The whiskers show the extremal measurements. (\textbf{C}) A $T_{2,Echo}$ measurement of Device 18a, fit with a stretched exponential. The fit gives $T_{2,Echo}$ = 249 $\pm$ 4 $\mu$s. (\textbf{D}) $T_{2,CPMG}$ of Device 11c as a function of the number of gates in a CPMG pulse sequence. }
\end{figure}

\begin{figure}[!t]
	\begin{center}
		\includegraphics[width=4.75in]{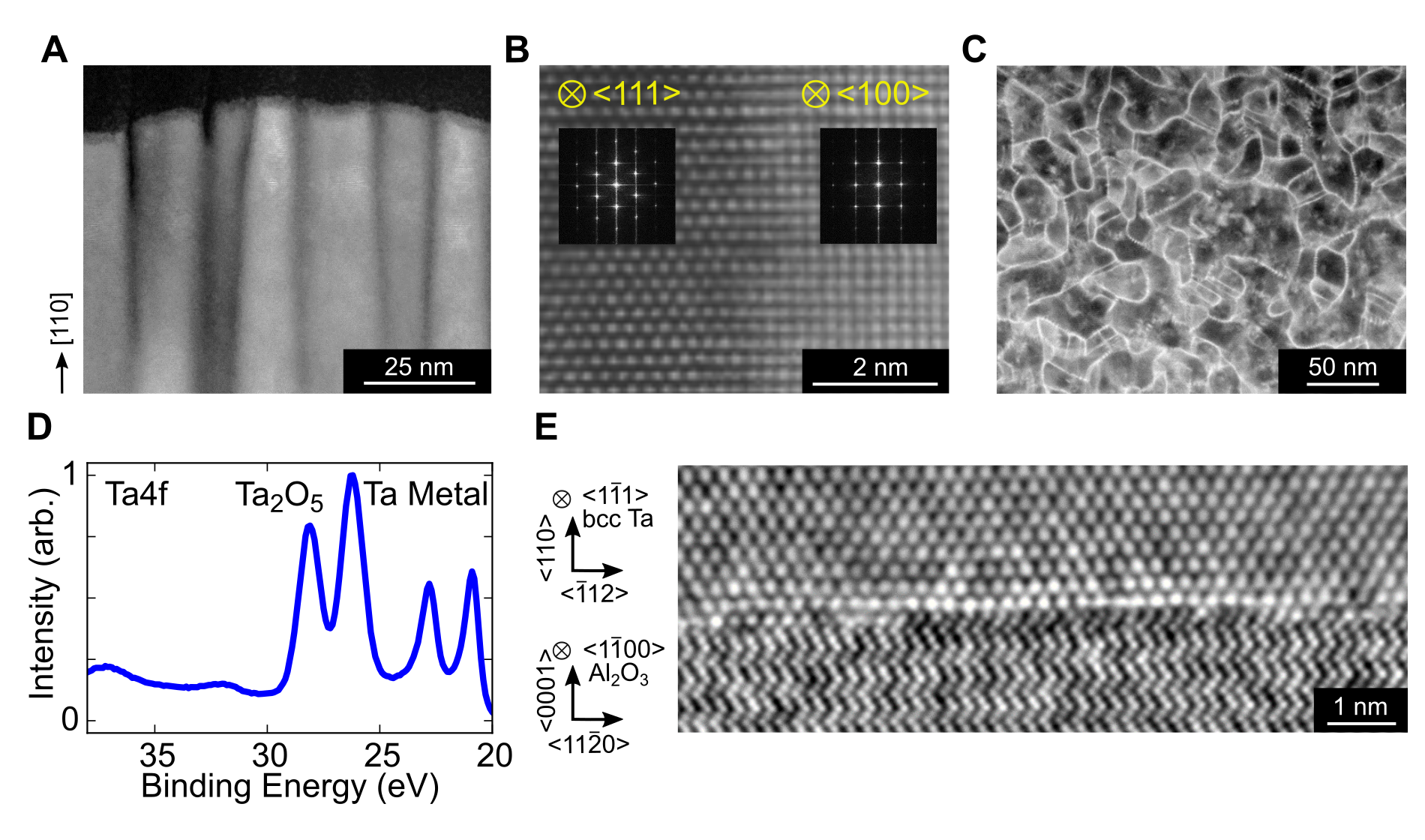}
	\end{center}
	\vspace{-0.6cm}
	\caption*{\textbf{Fig. 3. Microscopy and spectroscopy of tantalum films.} ($\textbf{A}$) STEM image of the tantalum film, showing single-crystal columns with the growth direction oriented along the [110] axis. ($\textbf{B}$) Atomic resolution STEM image of an interface between two columns, viewed from $\braket{1\bar{1}1}$ and $\braket{001}$ zone axes respectively. Fourier transforms (insets) of the image show that the columns are oriented with the image plane perpendicular to the $\braket{111}$ or $\braket{100}$  directions. ($\textbf{C}$) STEM image of a horizontal device cross section, showing grain boundaries. Image contrast at grain boundaries results from diffraction contrast caused by interfacial defects \cite{materialsAndMethods}. ($\textbf{D}$) XPS spectrum of a device, exhibiting peaks from tantalum metal and Ta$_2$O$_5$ \cite{mcguireCoreBETa}. (\textbf{E}) High-resolution STEM with integrated differential phase contrast imaging of the interface between the sapphire and tantalum showing epitaxial growth.}
\end{figure}

% Supplemental Information Figures

\begin{figure}[!t]
	\begin{center}
		\includegraphics[width=4.75in]{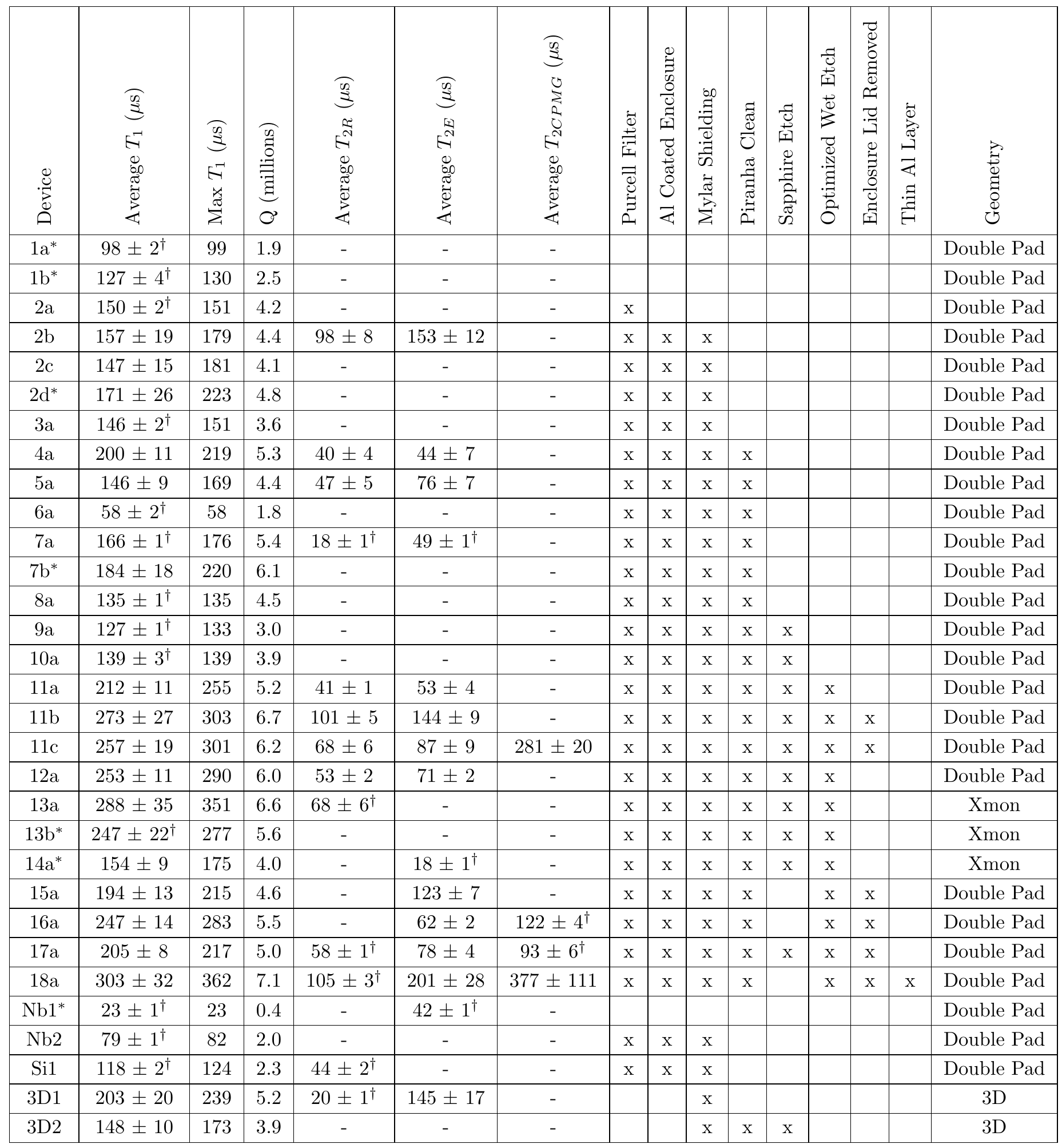}
	\end{center}
	\vspace{-0.6cm}
	\caption*{\textbf{Table S1. Summary of devices.} Here we include measurements of devices with different designs, fabrication procedures, and packaging. Devices labeled "Nb" were made with niobium instead of tantalum (Nb1 was heated to 350$^\circ$C then cooled for 20 minutes before deposition, Nb2 was deposited at approximately 500$^\circ$C) and all other devices were made from tantalum. Device Si1 was composed of about 200 nm of tantalum deposited on high-resistivity silicon. Devices labeled with the same number but different letters indicate the same qubit measured in different measurement cycles. Entries marked with a "$\dagger$" had three or fewer repeated measurements, and the reported errors were calculated by propagating the fit uncertainties. Otherwise the errors were calculated by finding the standard deviation of multiple measurements. Devices labeled with a "$\ast$" were fit without constraining the line of best fit to be normalized and have the proper offset. The average $T_{2,CPMG}$ column denotes the time averaged dynamical decoupling decoherence time at an optimal gate number.} 
\end{figure}

\begin{figure}[!t]
	\begin{center}
		\includegraphics[width=2.25in]{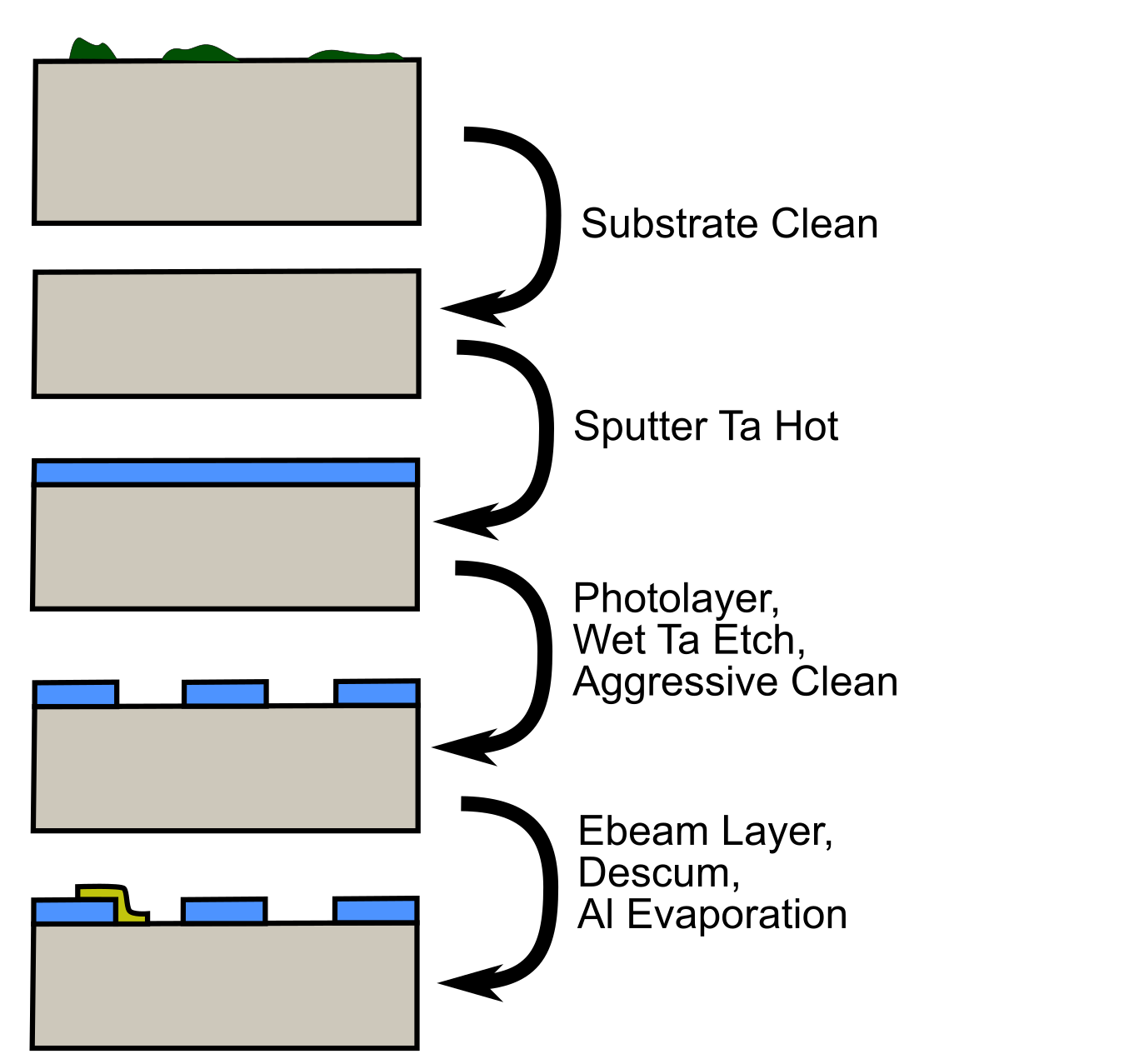}
	\end{center}
	\vspace{-0.6cm}
	\caption*{\textbf{Fig. S1. Qubit fabrication process.} The sapphire substrate (gray) is initially contaminated with carbon (green) which is reduced through substrate cleaning.  Tantalum (blue) is then deposited and subsequently patterned with a wet etch. Finally, the Josephson junctions (yellow) are lithographically defined and deposited.} 
\end{figure}

\begin{figure}[!t]
	\begin{center}
		\includegraphics[width=4.75in]{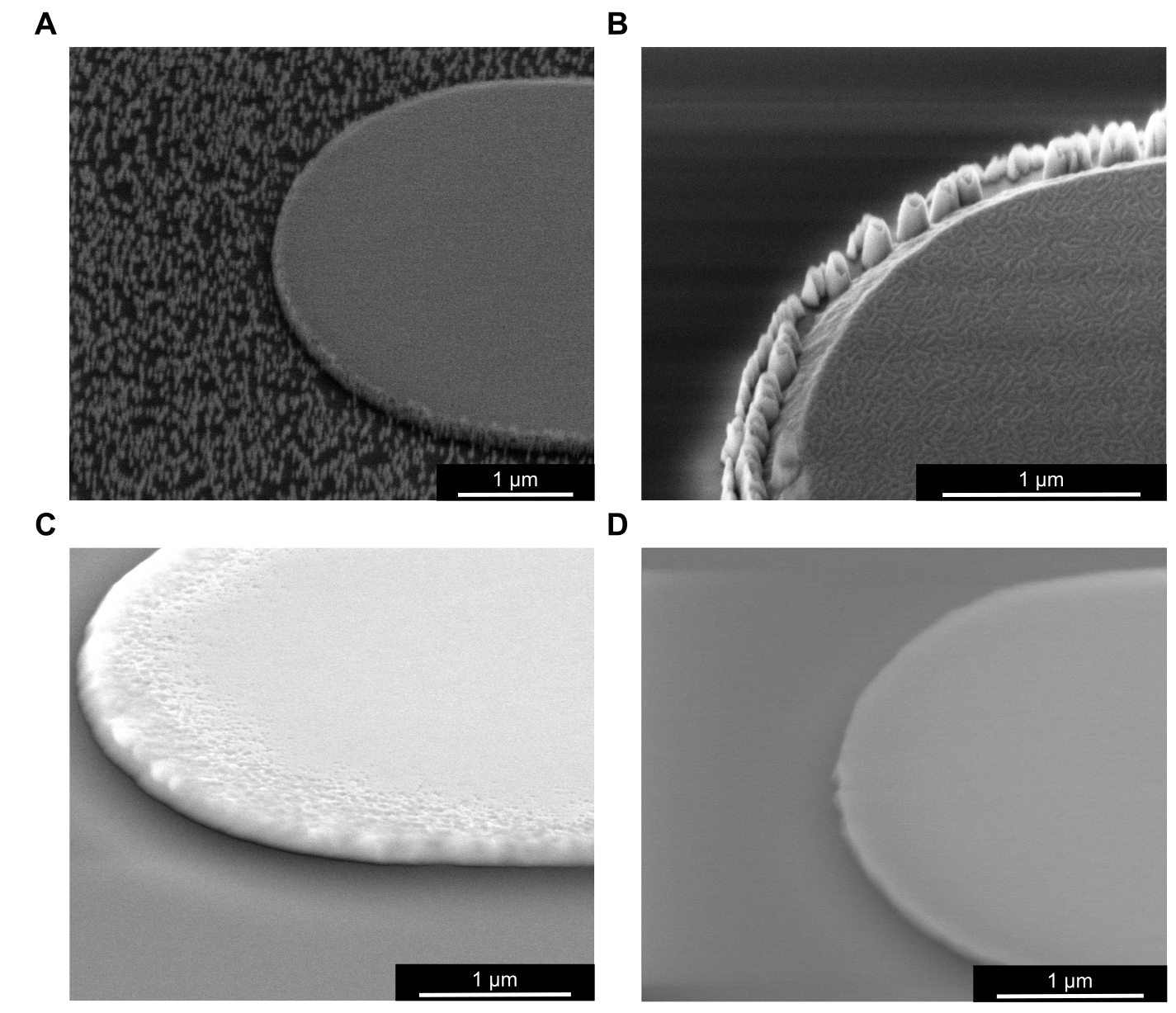}
	\end{center}
	\vspace{-0.6cm}
	\caption*{\textbf{Fig. S2. Scanning electron microscopy images of tantalum etch development.} In all panels tantalum capacitor pads on a sapphire substrate protrude from the right side of the image. (\textbf{A}, \textbf{B}) Examples of surface roughening after a 8:3:2 CHF$_3$:SF$_6$:Ar dry etch with a 5-7 mTorr pressure and RF/ICP power 30 W/30 W (\textbf{A}) and 100 W/100 W (\textbf{B}). The rough features near the sidewalls in (\textbf{B}) survived both a piranha etch and an oxygen plasma etch. (\textbf{C}) Initial wet etch results showed roughening of the tantalum near the edge of the pad, which was circumvented (\textbf{D}) by employing a thicker photoresist.
 } 
\end{figure}

\begin{figure}[!t]
	\begin{center}
		\includegraphics[width=2.25in]{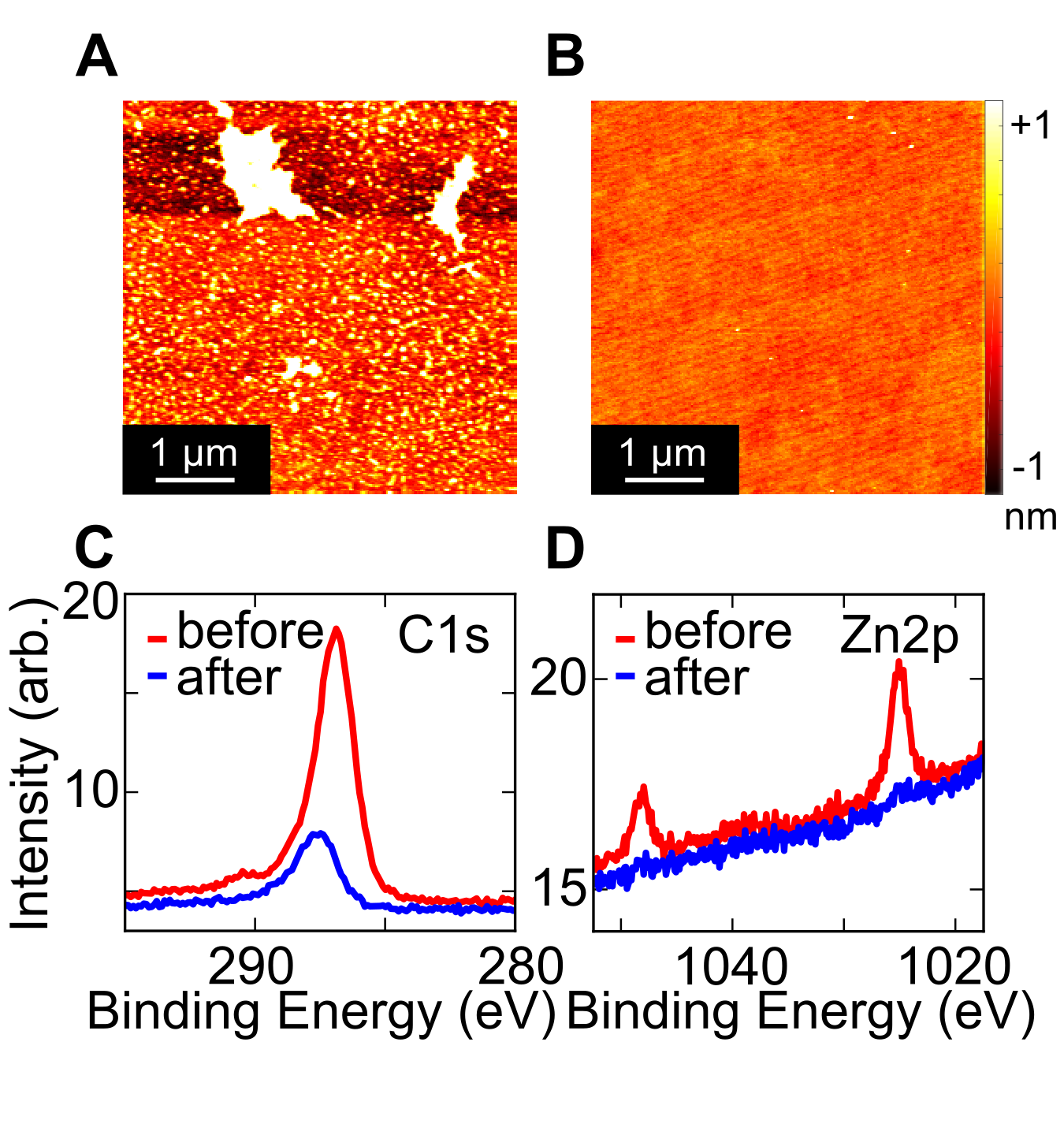}
	\end{center}
	\vspace{-0.6cm}
	\caption*{\textbf{Fig. S3. Characterization of sapphire surface.} AFM images of sapphire after dicing, stripping resist, and solvent cleaning (\textbf{A}) and after subsequent piranha cleaning and etching (\textbf{B}), showing the removal of particulates from the surface. XPS of sapphire identifies carbon (\textbf{C}) and zinc (\textbf{D}) contaminants on the sapphire surface. After piranha cleaning and etching, carbon is reduced by around a factor of five, and zinc is no longer detected. "Before" corresponds to the surface after dicing and solvent cleaning but before acid procedures, and "after" is following acid cleaning steps.} 
\end{figure}

\begin{figure}[!t]
	\begin{center}
		\includegraphics[width=4.75in]{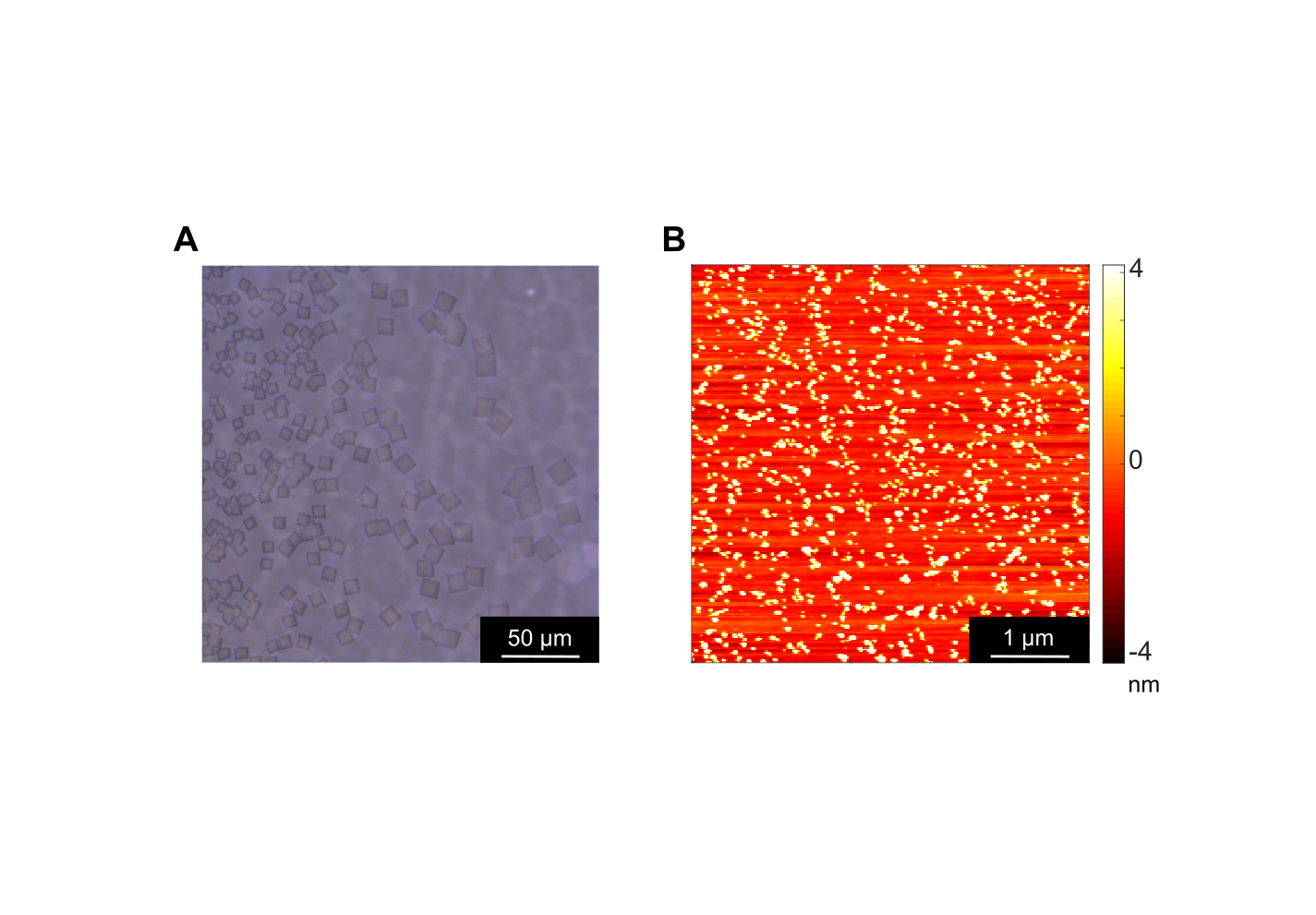}
	\end{center}
	\vspace{-0.6cm}
	\caption*{\textbf{Fig. S4. Sapphire processing pitfalls.} ($\textbf{A}$) Optical microscope image of crystals on sapphire after etching in refluxing sulfuric acid for 30 min. ($\textbf{B}$) AFM image of sapphire surface showing particulate contaminants after etching and piranha cleaning in borosilicate glassware} 
\end{figure}

\begin{figure}[!t]
	\begin{center}
		\includegraphics[width=4.75in]{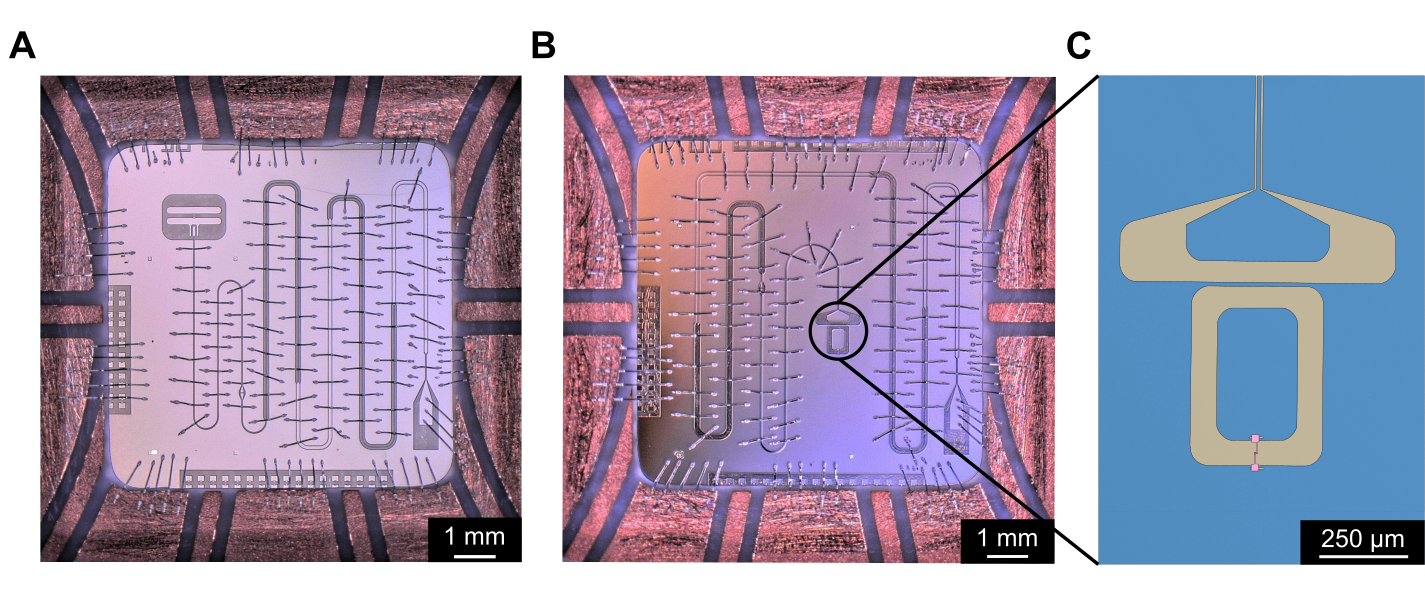}
	\end{center}
	\vspace{-0.6cm}
	\caption*{\textbf{Fig. S5. Device geometry overview.} Double-pad (\textbf{A}) and Xmon geometry (\textbf{B}) transmons mounted to a PCB. In each case, the chip is pressed beneath an opening in a PCB that has copper traces, here visible around the outside of the images. The excitation and measurement pulses first enter the curving Purcell filter, go through a capacitive coupler to the resonator, then to the qubit. We note in (\textbf{B}) that we moved the qubit to the center of the chip. (\textbf{C}) Close-up, false-colored SEM image of the Xmon qubit and coupler. 
 } 
\end{figure}

\begin{figure}[!t]
	\begin{center}
		\includegraphics[width=4.75in]{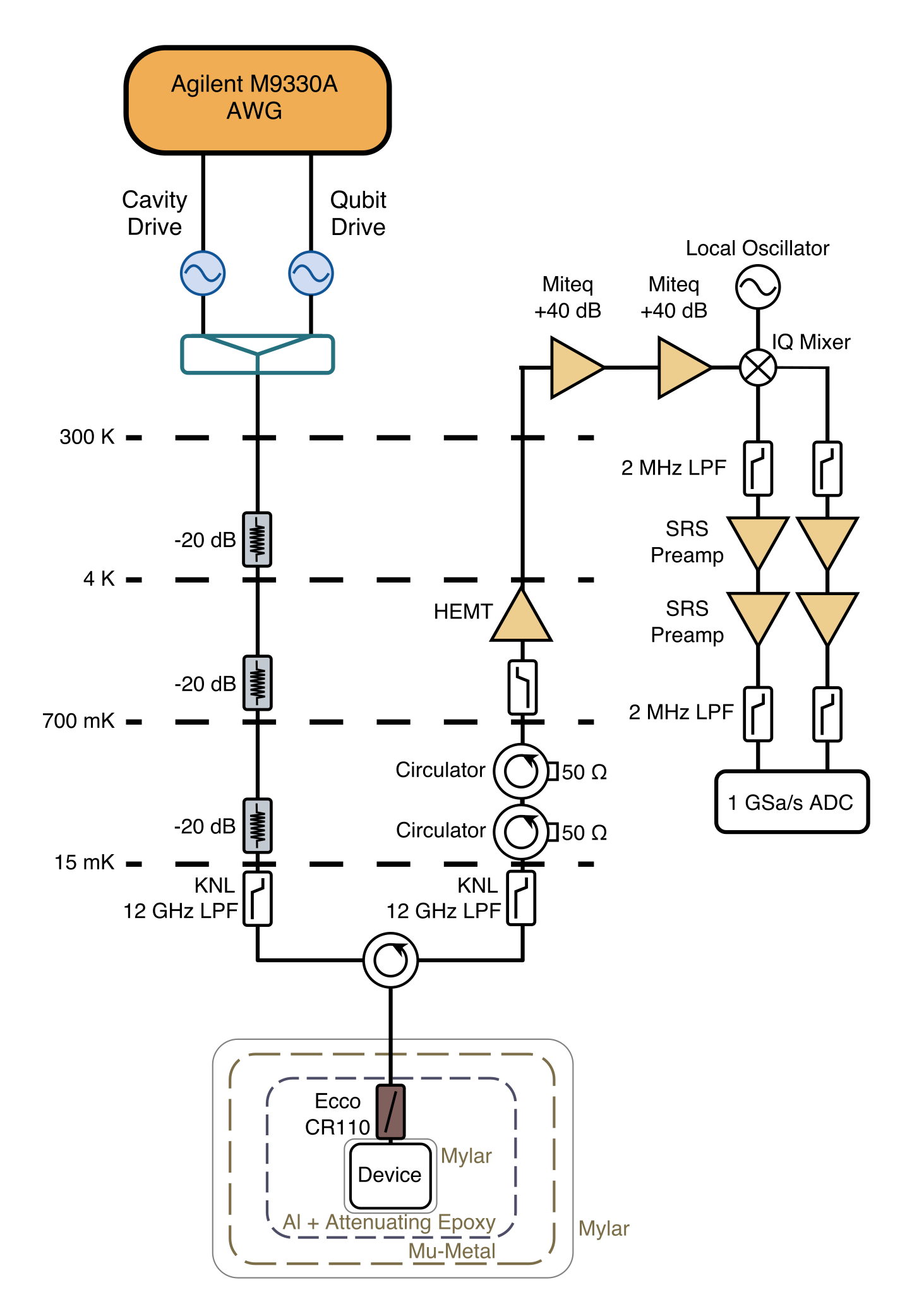}
	\end{center}
	\vspace{-0.6cm}
	\caption*{\textbf{Fig. S6. Schematic of the measurement electronics and device shielding.}
 } 
\end{figure}

\begin{figure}[!t]
	\begin{center}
		\includegraphics[width=2.5in]{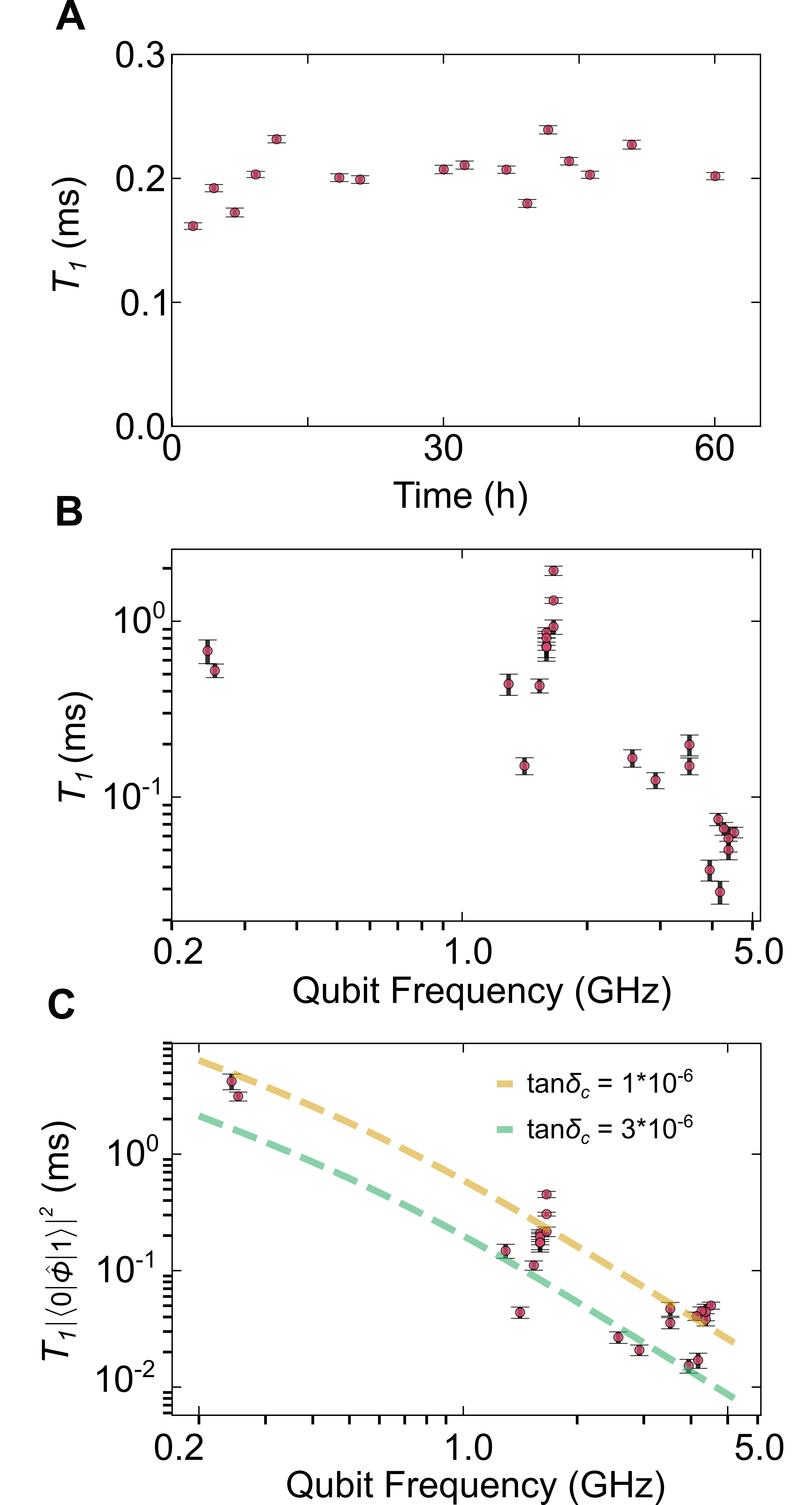}
	\end{center}
	\vspace{-0.6cm}
	\caption*{\textbf{Fig. S7. Tantalum 3D and fluxonium devices.} (\textbf{A}) Measurement of $T_1$ over time for Device 3D1. (\textbf{B}) Fluxonium $T_1$ as a function of frequency which can be fit to determine the dielectric loss tangent (\textbf{C}) when combined with the phase matrix element between logical qubit states, $\langle0|\hat{\phi}|1\rangle$.
 } 
\end{figure}

\begin{figure}[!t]
	\begin{center}
		\includegraphics[width=4.75in]{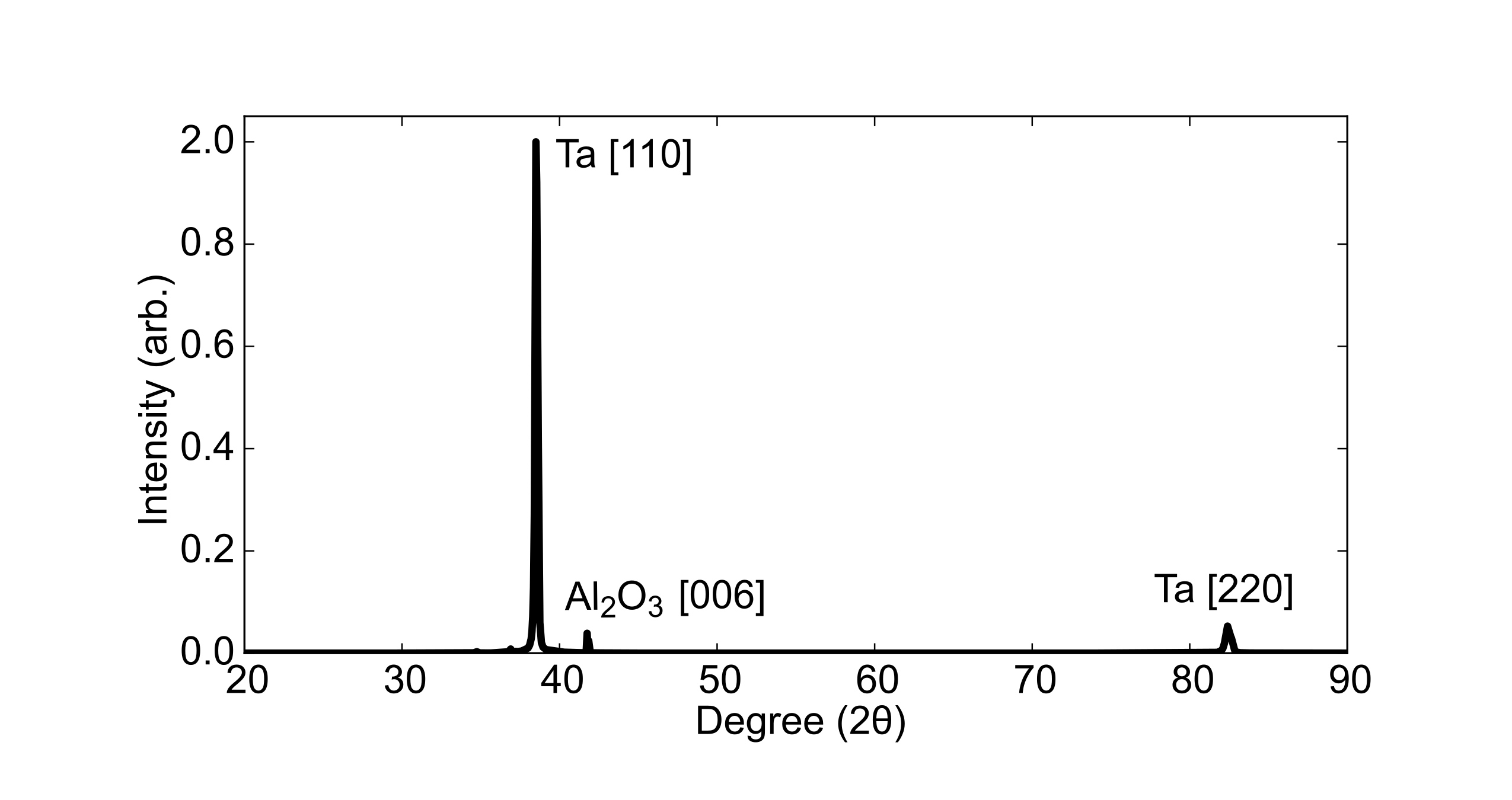}
	\end{center}
	\vspace{-0.6cm}
	\caption*{\textbf{Fig. S8.  X-ray diffraction spectrum of a sputtered tantalum film on sapphire.} XRD spectrum of sputtered tantalum on sapphire shows clear peaks corresponding to $\alpha$-tantalum and sapphire. A few unassigned small peaks are also visible which could be caused by contamination, instrumental artifacts, or impurities in the films. } 
\end{figure}

\begin{figure}[!t]
	\begin{center}
		\includegraphics[width=4.75in]{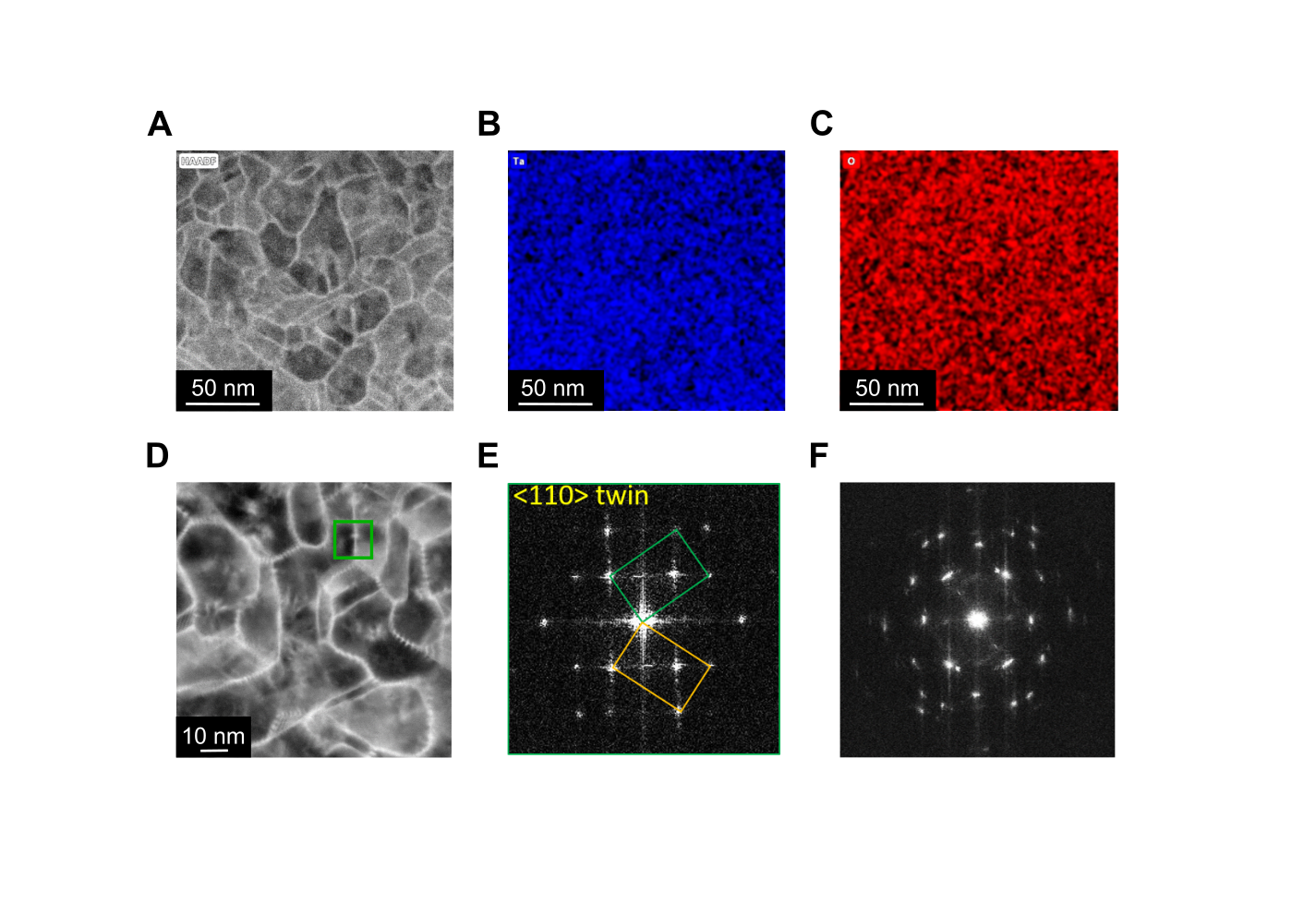}
	\end{center}
	\vspace{-0.6cm}
	\caption*{\textbf{Fig. S9. Grain boundary characterization.} ($\textbf{A}$) Plane-view STEM image showing grain boundaries. ($\textbf{B, C}$) EDS images of the same region shown in (\textbf{A}), displaying a uniform distribution of tantalum ($\textbf{B}$) and oxygen ($\textbf{C}$).  ($\textbf{D}$) Atomic resolution STEM image of the boundaries. ($\textbf{E}$) Fourier transform of the STEM image at a grain boundary indicated by the green box region of ($\textbf{D}$), showing a pattern consistent with twinning.  ($\textbf{F}$) Fourier transform of the entire image in ($\textbf{D}$) shows the rotational symmetries of the grains. } 
\end{figure}

\begin{figure}[!t]
	\begin{center}
		\includegraphics[width=4.75in]{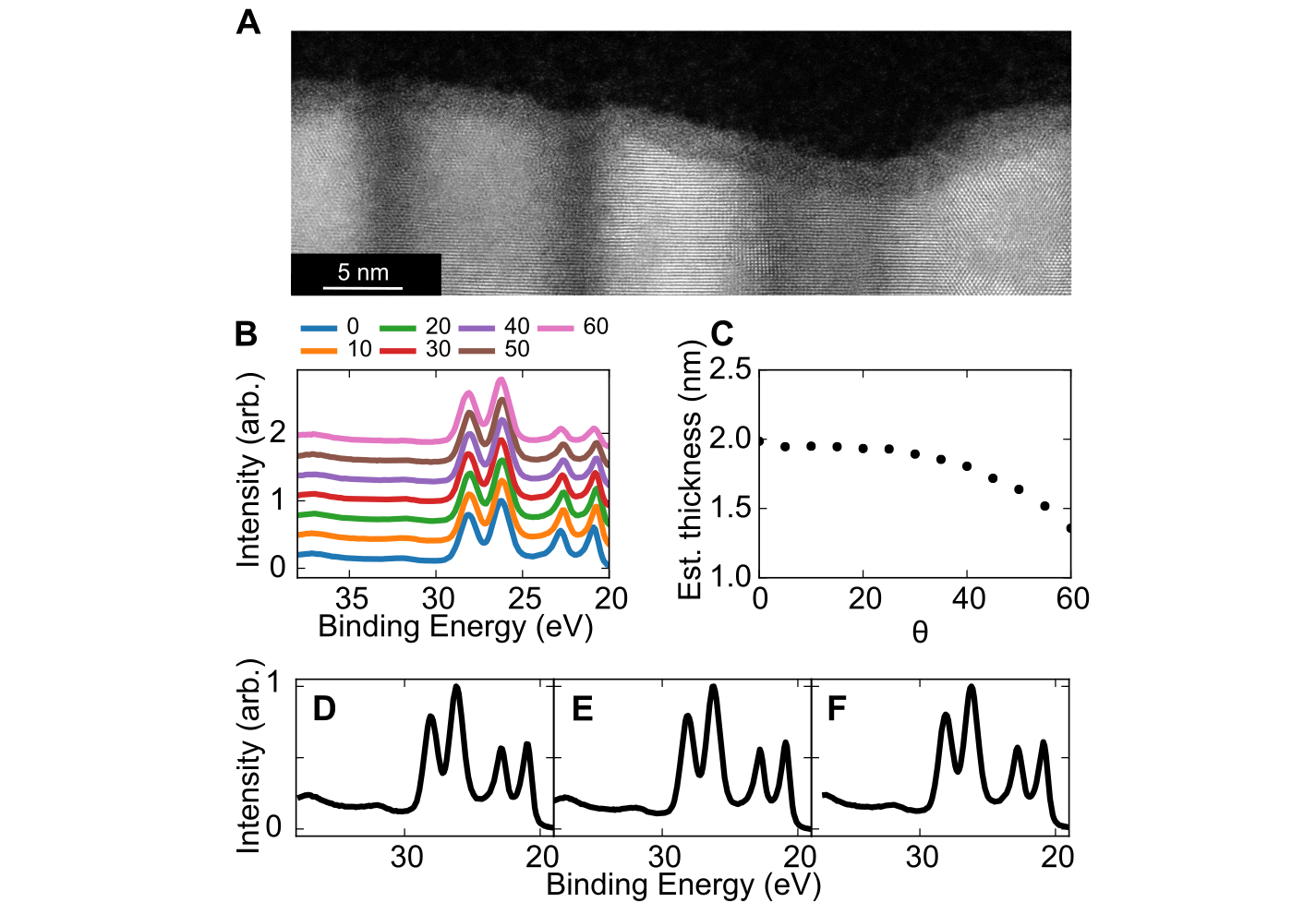}
	\end{center}
	\vspace{-0.6cm}
	\caption*{\textbf{Fig. S10. Oxide characterization.} ($\textbf{A}$)	Atomic resolution STEM image showing an amorphous oxide layer about 2-3 nm thick on the tantalum surface. ($\textbf{B}$) Angle-resolved XPS measurements of Ta4f region of a fabricated device, offset vertically for clarity. Colors indicate the angle in degrees between sample and detector. ($\textbf{C}$) Estimated oxide thickness as a function of angle between sample and detector. ($\textbf{D, E, F}$) Ta4f normal incidence XPS data of three completed devices showing nearly identical spectra. The devices were from different tantalum depositions and underwent different fabrication steps. In addition to other variations in fabrication, the device in ($\textbf{D}$) was only solvent cleaned while the devices surveyed in ($\textbf{E}$) and ($\textbf{F}$) were piranha cleaned. 
 } 
\end{figure}

\begin{figure}[!t]
	\begin{center}
		\includegraphics[width=2.25in]{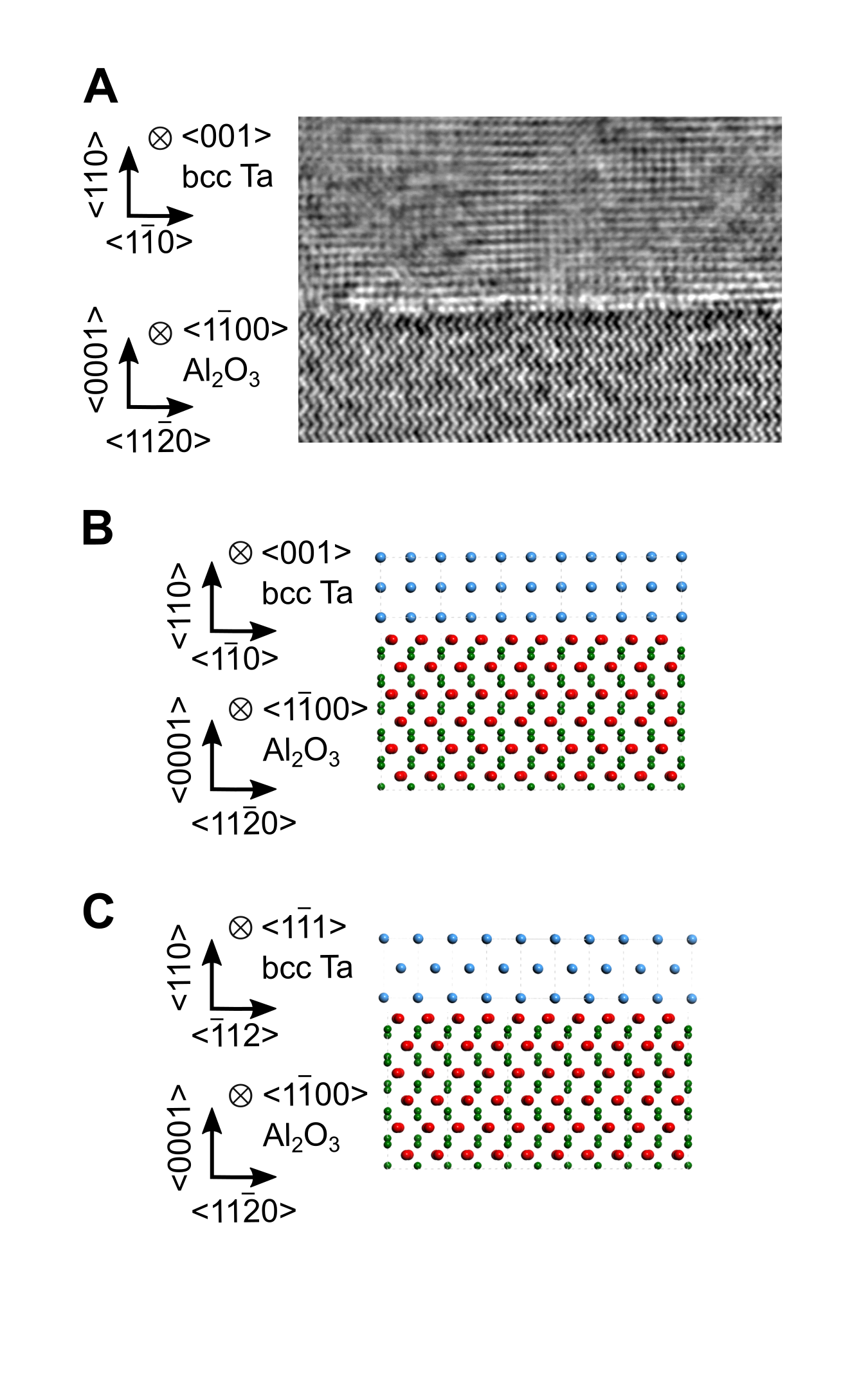}
	\end{center}
	\vspace{-0.6cm}
	\caption*{\textbf{Fig. S11. Tantalum-sapphire interface characterization.} ($\textbf{A}$) Atomic resolution iDPC STEM image showing the interface between tantalum and sapphire with the image plane perpendicular to the $\braket{100}$ direction of tantalum. ($\textbf{B}$) Atomistic model of the ideal interface for the tantalum column orientation shown in (\textbf{A}). ($\textbf{C}$) Atomistic model of the ideal interface for the tantalum column orientation shown in Fig. 3E. In both cases oxygen atoms are depicted in red, aluminum in green, and tantalum in blue. 	 
 } 
\end{figure}

\begin{figure}[!t]
	\begin{center}
		\includegraphics[width=4.75in]{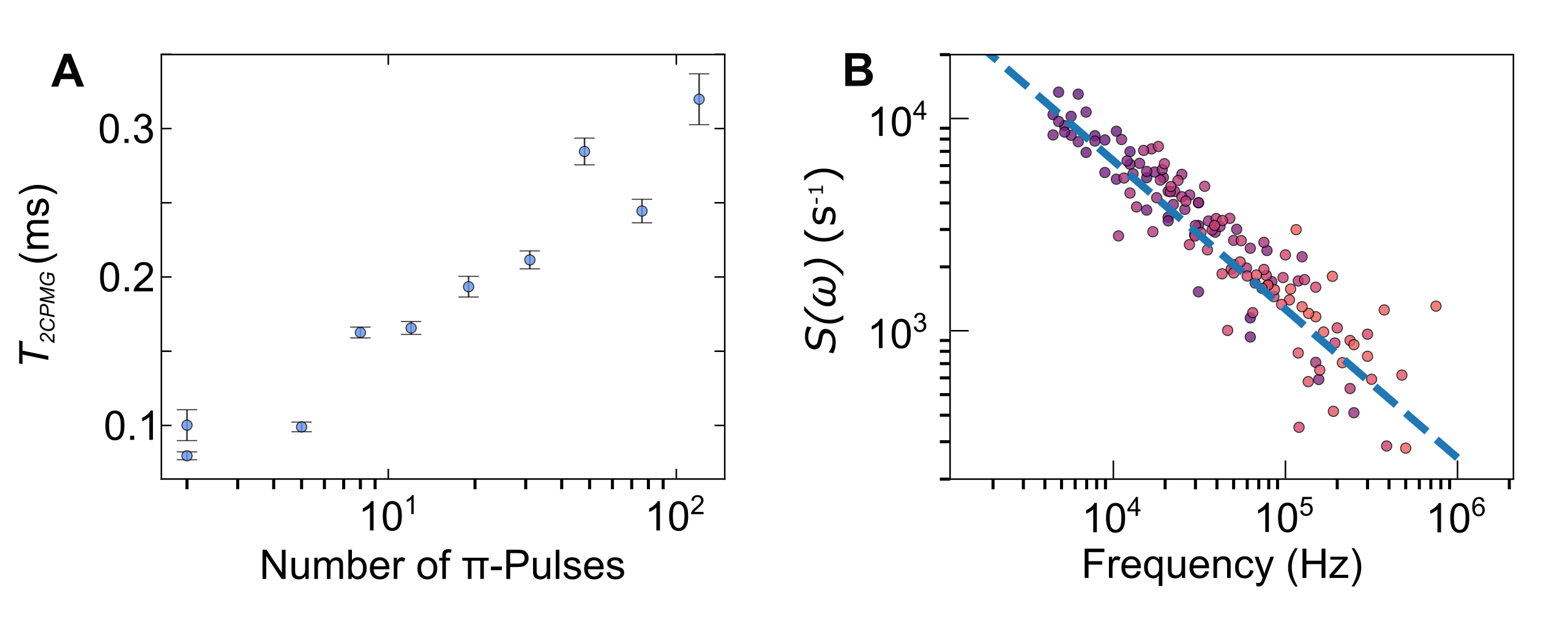}
	\end{center}
	\vspace{-0.6cm}
	\caption*{\textbf{Fig. S12. CPMG noise spectrum of Device 11c.} ($\textbf{A}$) $T_{2,CPMG}$ as an increasing number of pulses reduce the qubit's sensitivity to low-frequency noise. ($\textbf{B}$) Noise power spectral density, following \cite{bylander2011noise}. The red dashed line indicates a fit by eye to $A/f^\alpha + B$ where $\alpha$ = 0.7, A = 2e6$s^{-1}$, and B = 3e2$s^{-1}$.} 
\end{figure}

\begin{figure}[!t]
	\begin{center}
		\includegraphics[width=2.25in]{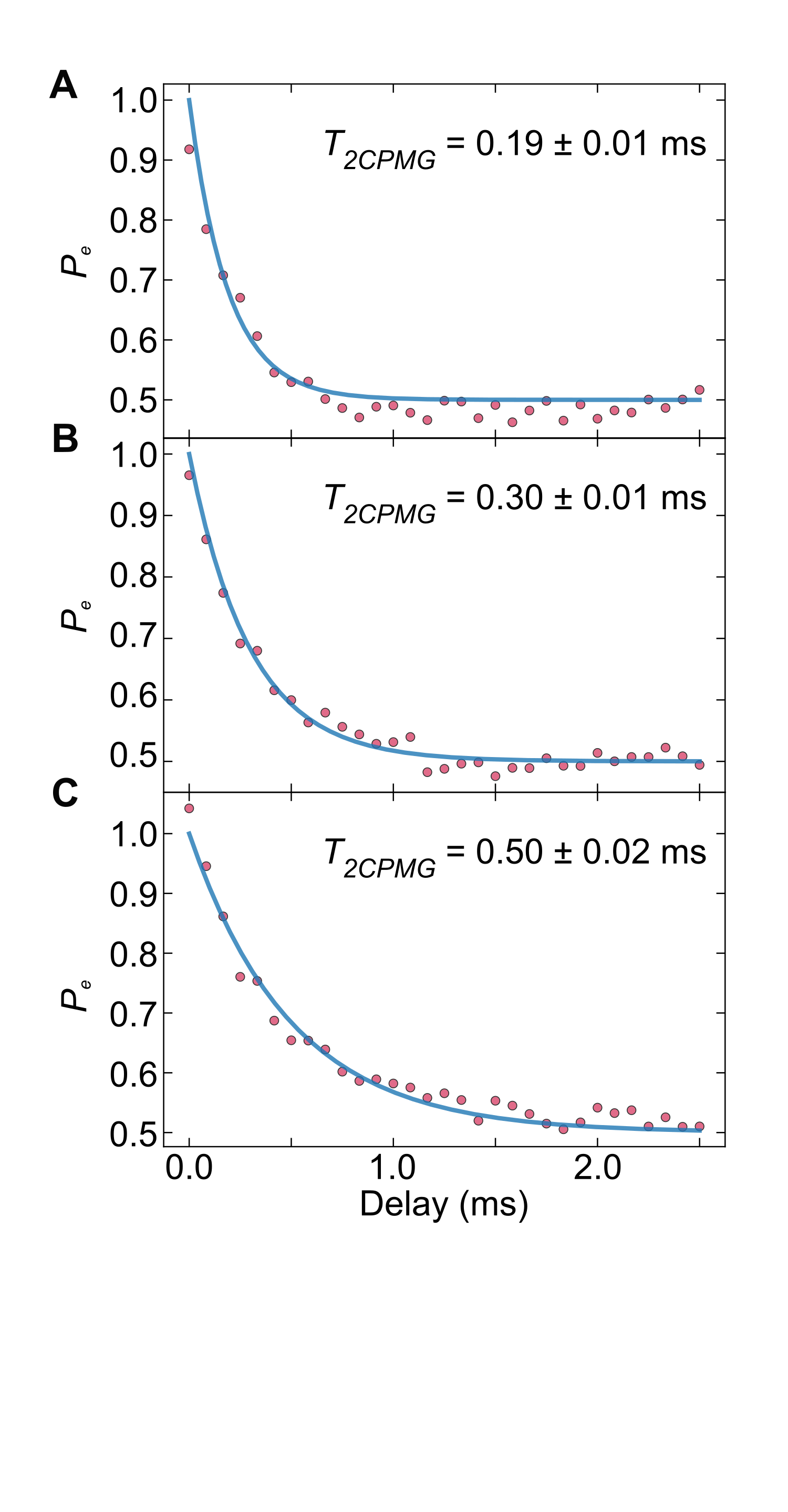}
	\end{center}
	\vspace{-0.6cm}
	\caption*{\textbf{Fig. S13. CPMG traces.} Low (\textbf{A}), middle (\textbf{B}), and high (\textbf{C}) $T_{2,CPMG}$ traces from the data in Fig. 2A. All three traces were fit to a stretched exponential with the exponent constrained to be larger than one.}
\end{figure}

\end{document}